\documentclass[10pt,journal,twocolumn]{IEEEtran}
\usepackage{wrapfig,booktabs,fancyhdr,amsmath,amsfonts,tabularx,numprint,dsfont}
\usepackage{cite,bm,bbm,amssymb,amsthm,url,multirow,times,enumitem}
\usepackage{mathtools,siunitx,balance,tikz,adjustbox,array,graphicx,xspace}
\usepackage{mathrsfs}
\usepackage[font=footnotesize]{subcaption}
\usepackage[linesnumbered,ruled]{algorithm2e}
\usepackage[colorlinks=true, linkcolor=black, citecolor=blue, urlcolor=blue]{hyperref}
\usepackage{tikz}

 \usetikzlibrary{decorations.pathreplacing}
 \usetikzlibrary{backgrounds}
 \pgfdeclarelayer{nodelayer}
 \pgfdeclarelayer{edgelayer}
 \pgfsetlayers{nodelayer,edgelayer,main}
 \usetikzlibrary{arrows.meta}
 \usepackage{svg}
 \usepackage{float}

\tikzstyle{none_node}=[font={\normalsize}, inner sep=0pt, minimum size=0pt, align=center]
\tikzstyle{rect}=[fill=white, draw=black, shape=rectangle, rounded corners=1, align=center]
\tikzstyle{box}=[fill=white, draw=black, shape=rectangle, align=center, minimum height=4cm, minimum width=4cm, font={\huge}]
\tikzstyle{box_smaller}=[fill=white, draw=black, shape=rectangle, align=center, minimum height=4cm, minimum width=4cm, font={\large}]

\tikzstyle{none}=[draw=none, fill=none]

\tikzstyle{rightarrow}=[->, line width=1pt, join=round]
\tikzstyle{dashed_ln}=[-, dashed, line width=1pt, join=round]
\tikzstyle{dashed_ln_rightarrow}=[->, dashed, line width=1pt, join=round]

\newcommand{\vb}{\boldsymbol}

\newcommand{\sfrac}[2]{\textstyle \frac{#1}{#2}}
\newcommand{\ba}{\begin{array}}
\newcommand{\ea}{\end{array}}
\newcommand{\sinc}{\text{sinc}}
\newcommand{\define}{\triangleq}

\DeclareMathAlphabet{\mathpzc}{OT1}{pzc}{m}{it}

\newcommand{\Fs}{\ensuremath{F_\text{s}}}

\newcommand{\N}{\ensuremath{N}}

\newcommand{\Tsym}{\ensuremath{T_\text{sym}}}

\newcommand{\Tf}{\ensuremath{T_\text{f}}}
\newcommand{\tf}{\ensuremath{t_\text{f}}}
\newcommand{\ts}{t_\text{s}}

\newcommand{\dtf}{\ensuremath{{\delta t}_\text{f}}}
\newcommand{\dts}{\ensuremath{{\delta t}_\text{s}}}
\newcommand{\dtc}{\ensuremath{{\delta t}_\text{c}}}

\newcommand{\dtfb}{\ensuremath{\bar{\delta t}_\text{f}}}
\newcommand{\tc}{\ensuremath{t_\text{c}}}
\newcommand{\td}{\ensuremath{t_\text{d}}}

\newcommand{\tr}{\ensuremath{t_\text{r}}}
\newcommand{\tu}{\ensuremath{t_\text{u}}}
\newcommand{\wwr}{\ensuremath{w_\text{r}}}
\newcommand{\wwu}{\ensuremath{w_\text{u}}}

\newcommand{\ttr}{\ensuremath{\tilde{t}_\text{r}}}
\newcommand{\ttu}{\ensuremath{\tilde{t}_\text{u}}}

\newcommand{\Mil}{\ensuremath{\mathcal{M}_{il}}}

\newcommand{\dTOF}{\ensuremath{\delta t_\text{tof}}}
\newcommand{\hatdTOF}{\ensuremath{\delta \hat{t}_\text{tof}}}

\newcommand{\T}{^{\mathsf{T}}}
\newcommand{\F}{\ensuremath{F}}
\newcommand{\Tg}{\ensuremath{T_\text{g}}}
\newcommand{\Ng}{\ensuremath{N_\text{g}}}
\newcommand{\Na}{\ensuremath{N_\text{a}}}
\newcommand{\Nag}{\ensuremath{N_\text{ag}}}
\newcommand{\Nbi}{\ensuremath{N_{\text{b}i}}}

\newcommand{\Nasi}{\ensuremath{N_{\text{as}i}}}
\newcommand{\Tag}{\ensuremath{T_\text{ag}}}

\newcommand{\Ff}{\ensuremath{F_\text{f}}}
\newcommand{\Tfg}{\ensuremath{T_\text{fg}}}

\DeclareSIUnit{\dBi}{dBi}

\begin{document}
\title{Toward Pseudorange-Based PNT with Starlink: \\ A Mock Fused LEO GNSS Implementation}

\author{
  \IEEEauthorblockN{Samuel C. Morgan\IEEEauthorrefmark{1}, Zacharias M. Komodromos\IEEEauthorrefmark{2}, Wenkai Qin\IEEEauthorrefmark{2}, \\ Andrew Graff\IEEEauthorrefmark{3}, Zachary L. Clements\IEEEauthorrefmark{1}, W. Jeremy Morrison\IEEEauthorrefmark{1}, Todd E. Humphreys\IEEEauthorrefmark{1}} \\
  \IEEEauthorblockA{\IEEEauthorrefmark{1}\textit{Department of Aerospace
      Engineering and Engineering Mechanics, The University of Texas at Austin}} \\
  \IEEEauthorblockA{\IEEEauthorrefmark{2}\textit{Department of Electrical and Computer Engineering, The University of Texas at Austin}} \\
  \IEEEauthorblockA{\IEEEauthorrefmark{3}\textit{CesiumAstro}} \\
}

\maketitle

\begin{abstract}
  We present experimental results from Starlink for pseudorange-based
  positioning, navigation, and timing (PNT), emulating a system that fuses a
  primary communications mission with a secondary low-Earth orbit (LEO) PNT
  mission. Legacy global navigation satellite system (GNSS) constellations face
  inherent vulnerabilities and signal limitations, prompting growing interest in
  broadband LEO constellations---such as SpaceX's Starlink---for resilient PNT.
  However, opportunistic use of stand-alone Starlink signals fails to achieve
  GNSS-competitive accuracy due to frame timing irregularities and reliance on
  uncorrected poor-quality publicly available
  ephemeris data. To overcome these limitations, we present a mock
  implementation of a fused LEO GNSS architecture. Our method extracts
  time-of-arrival observables from live Starlink Ku-band downlinks and pairs
  them with high-fidelity, beam-specific orbit and clock models distributed as
  if provided natively by the constellation operator. In a field campaign, and
  despite constraints imposed by consumer-grade receiver hardware, the method
  yields 95\% error bounds of 2.33~m horizontally (semi-major axis), 3.37~m
  vertically, and 11.19~ns in timing, with the actual solution errors falling
  well within these bounds.
\end{abstract}

\newif\ifpreprint
\preprintfalse

\ifpreprint
  \pagestyle{plain}
  \thispagestyle{fancy}
  \fancyhf{}
  \renewcommand{\headrulewidth}{0pt}
  \rfoot{\footnotesize \bf October 2020 preprint of paper submitted for review}
  \lfoot{\footnotesize \bf
    Copyright \copyright~2020 by Lakhsay Narula \\ and Todd E. Humphreys}

\else

  \thispagestyle{empty}
  \pagestyle{plain}

\fi

\section{Introduction}

High-rate data connectivity has become a foundational enabler of modern
services, including video streaming, distributed work, and low-latency extended
and augmented reality. Until recently, such capabilities were largely confined
to regions with established communications infrastructure or substantial
investment in new deployment. The rise of low Earth orbit (LEO) broadband
mega-constellations has fundamentally altered this landscape by enabling global,
infrastructure-independent coverage.

Achieving high throughput for many simultaneous users requires wide signal
bandwidths. These bandwidths, in turn, enable highly precise time-of-arrival
(TOA) measurements, which form the fundamental observable underlying
positioning, navigation, and timing (PNT) in global navigation satellite systems
(GNSS). In principle, therefore, broadband LEO constellations are well suited to
provide PNT services. In practice, however, existing systems (e.g., SpaceX's
Starlink and Eutelsat's OneWeb) have not translated this capability into
user-facing PNT. Starlink, for example, employs an internal positioning mode to
support terminal network entry and, until recently, offered users access to its
position estimates, whose accuracy had improved to roughly \SI{20}{\meter};
SpaceX has since withdrawn this access~\cite{goward2026starlinklocation}.

As a largely invisible utility, GNSS underpin critical sectors ranging from
agriculture and aviation to telecommunications and financial markets. No other
technology currently delivers the globally available meter-level positioning and
nanosecond-level timing on which safe transportation, synchronized networks, and
precise transaction time-stamping depend. Yet legacy civil GNSS were not
designed for adversarial environments. Their weak L-band signals, transmitted
from medium Earth orbit and broadcast as open-access services, leave users
vulnerable to jamming and spoofing. These risks are no longer hypothetical; they
are intensifying alongside the growing role of electronic warfare, causing
predictable and escalating consequences for civilian users of GNSS-based
PNT~\cite{clements2023plansdirectgeo,clements2026spoofergeo,humphreysGNSShandbook,psiakinewbluebookspoofing}.

These vulnerabilities motivate the search for complementary or alternative PNT
architectures capable of delivering accuracy and resilience under degraded GNSS
conditions. LEO constellations are particularly attractive in this role. Their
low-altitude orbits yield higher received signal power than medium Earth orbit
GNSS, and their rapid motion across the sky appreciably changes the
user-satellite geometry within tens of seconds. The higher power permits more
precise pseudorange measurements, while the changing geometry improves position
observability and causes spatial errors, such as multipath, to decorrelate
quickly. LEO's shorter range also makes it easier to close a two-way link
between user and satellite, which is significant since two-way signaling is the
only way to fully secure PNT~\cite{narula2018securesynchronization}.
Consequently, LEO-based PNT has emerged as a promising candidate for enhancing
the robustness of civil PNT infrastructure~\cite{reid2018broadbandleo,
ProlF.S.2022PNaT, stock2025survey, hui2025review}, with the potential to
complement or backstop GNSS in the presence of jamming and
spoofing~\cite{humphreysGNSShandbook, opsgroup2024spoofing,
clements2026spoofergeo, zulhusni2025gpsinterference}.

Existing efforts in LEO PNT can be broadly categorized into four approaches:
dedicated, fused, network-aided opportunistic, and stand-alone opportunistic.
Table~\ref{tab:leoPntTable} summarizes these approaches in terms of deployment
cost, availability, positioning accuracy, time to first fix, and dependence on
GNSS.
\begin{table*}[t]
  \centering
  \caption{LEO PNT trade space}
  \begin{tabular}[c]{lllll}
    \toprule
                          & Dedicated                                                                                                           & Fused                                                                   & Network-Aided                & Stand-Alone                 \\
                          &                                                                                                                     &                                                                         & Opportunistic                & Opportunistic               \\
                          & \cite{van2024technical,rubinov2024leopnt,miller2023snap}                                                            & \cite{iannucci2022fusedLeo,graff2024ziv,graff2025data,de2025integrated}

                          & \cite{zhou2025iridium, komodromos2025NetworkPlans,xu2025joint}

                          & \cite{tan2019positioning,psiaki2021navigation,neinavaie2021acquisition,mclemore2022navigation,jardak2024leveraging}                                                                                                                                        \\ \midrule
    Description           & LEO constellation                                                                                                   & Fuses a secondary PNT                                                   & Like stand-alone             & Exploits unmodified signals \\
                          & or hosted payloads                                                                                                  & service with a primary                                                  & opportunistic, but a network & from
    communications
    SVs.                                                                                                                                                                                                                                                                               \\
                          & dedicated solely                                                                                                    & communications service.
                          & of reference stations                                                                                               & Public ephemerides.                                                                                                                  \\
                          & to PNT.                                                                                                             &                                                                         & provides corrections.        & No network aiding.          \\ \midrule
    Marginal              & high: constellation                                                                                                 & low: uses communications                                                & high: network of             & minimal                     \\
    deployment cost       & of SVs, ground network                                                                                              & hardware and signals                                                    & reference stations           &                             \\ \midrule
    Potential             & mid-term                                                                                                            & near-term                                                               & near-term for                & immediate                   \\
    availability          &                                                                                                                     &                                                                         & local coverage               &                             \\ \midrule
    Potential             & $<1$                                                                                                                & $<1$                                                                    & $<1$                         & $<100$                      \\
    accuracy (m)          &                                                                                                                     &                                                                         &                              &                             \\ \midrule
    Time to first fix (s) & $<10$                                                                                                               & $<10$                                                                   & $<10$                        & $<1000$                     \\ \midrule
    Dependency on         & @SVs in near-term                                                                                                   & @SVs in near-term                                                       & @reference stations
                          & @SVs in some cases                                                                                                                                                                                                                                         \\
    legacy GNSS           &                                                                                                                     &                                                                         & in near-term                 &                             \\ \bottomrule
  \end{tabular}
  \label{tab:leoPntTable}
\end{table*}

Several studies explore dedicated LEO PNT systems~\cite{van2024technical,
rubinov2024leopnt}. The authors of~\cite{miller2023snap} developed one of the
first software-defined receivers for such a constellation---Xona Space Systems'
Pulsar. The fused LEO GNSS concept of operations was introduced
in~\cite{iannucci2022fusedLeo, iannucci2020fused}, while subsequent works
present methods for extracting high-quality TOA measurements from existing
communications infrastructure within this paradigm~\cite{graff2024ziv,
qin2025toaDopplerPlans, graff2025data}. To further support fused LEO PNT, a
modification to the 5G non-terrestrial network standard was proposed
in~\cite{de2025integrated}.
Network-aided opportunistic approaches exploit unmodified signals but rely on a
network of reference stations for corrections. A third-party network that
distributes clock and orbit models for OneWeb is envisioned
in~\cite{komodromos2025NetworkPlans}, achieving good positioning accuracy thanks
to OneWeb's wide bandwidth despite poor ephemerides and poor satellite vehicle
(SV) geometry. Stand-alone opportunistic approaches dispense with network aiding
altogether, relying solely on public ephemerides~\cite{tan2019positioning,
psiaki2021navigation, neinavaie2021acquisition, mclemore2022navigation,
jardak2024leveraging}.

The overwhelming majority of opportunistic LEO PNT methods, whether
network-aided or stand-alone, are Doppler-based~\cite{sallouha2025gasSurvey}.
Yet Doppler measurements cannot match pseudorange techniques in timing accuracy:
they achieve only millisecond-level precision compared to the nanosecond-level
precision of pseudorange, even under favorable
conditions~\cite{psiaki2021navigation, mclemore2022navigation,
baron2024implementation}. Because many applications of interest demand highly
precise timing, we focus here on pseudorange-based PNT derived from TOA
measurements.

Among current broadband providers, Starlink offers the strongest potential for
pseudorange-based PNT. Its constellation can direct up to 128 beams into a
single cell, yielding favorable geometry, and its \SI{240}{\mega\hertz} downlink
includes deterministic pilots that enable highly accurate TOA
measurements~\cite{qin2025starlinktimingproperties,
humphreys2023starlinkSignalStructure, komodromos2023IONsimulator,
qin2026starlinktemplate}. Even compared with Galileo’s \SI{51}{\mega\hertz}
AltBOC signal, Starlink provides far stronger multipath resistance without the
side-peak ambiguity problem~\cite{fan2023double}.

Nevertheless, stand-alone opportunistic pseudorange-based PNT with Starlink
remains intractable: the onboard clocks exhibit inconsistent, unmodelable
behavior~\cite{qin2025starlinktimingproperties}. Given that timing errors for
Starlink SVs are, in the worst case,
beam-specific~\cite{qin2025starlinktimingproperties}, a network-aided
opportunistic solution would seem impractical at scale, as reference stations
would need to be spaced closer than the \SI{20}{\kilo\meter} Starlink beam
diameter~\cite{pekhterev2021bandwidth}. By contrast, a fused
communications-and-PNT service where Starlink distributes real-time orbit and
clock models offers a feasible path forward~\cite{iannucci2022fusedLeo}. We
emulate such a service through a network-aided opportunistic setup: our
reference network provides the same models and pilot symbols that the
constellation operator would transmit. Thus, this network-aided opportunistic
architecture serves as a mock fused implementation.

In a prior conference version of this paper~\cite{morgan2025mockplans}, we
proposed treating Starlink as a fused LEO GNSS by defining a control segment
(CS) receiver and a user terminal (UT) receiver. Equipped with a high-gain
antenna and a high-quality clock, the CS derived orbit and clock models from its
observations and relayed them to the UT, emulating data that would have been
provided in a downlink transmission. The UT then applied these models to form
pseudorange measurements and compute a PNT solution. A field campaign
demonstrated this concept, achieving \SI{10}{\meter} PNT accuracy despite
limited bandwidth and residual ephemeris errors.

Our previous work had several limitations: (1) rather than forming local
replicas, we obtained TOAs by directly cross-correlating in-phase and quadrature
samples from the CS and UT, which is bandwidth-intensive and noisy; (2) we
derived Ziv--Zakai bound (ZZB) optimal pilot allocations (as
in~\cite{graff2024ziv}) but did not explore them experimentally; (3) we
estimated TOAs by nearest-sample correlation instead of by superior
super-resolution techniques; and (4) we relied on two-line element (TLE)
ephemerides. Together, these limitations constrained PNT accuracy to the
\SI{10}{\meter} level.

In this paper, we address these limitations and extend our previous work. Our
primary contributions are (i) noise-free local replicas formed at the UT from
decoded CS data, which permit evaluation of both full-frame correlations and
ZZB-optimal pilot allocations; (ii) super-resolution TOA estimation, which
refines TOA estimates well beyond sample-level resolution; (iii) the replacement
of TLEs with precise SpaceX-published ephemerides, which significantly reduces
orbit-related errors in the positioning solutions; and (iv) a comparison of the
positioning performance of ZZB-optimal pilot allocations against that of the
full frame, with both nearest-sample and super-resolution TOA estimation.

Assuming a static UT with an oven-controlled crystal oscillator (OCXO) and
super-resolution TOA estimation via full-frame correlation, the proposed
extensions yield \SI{95}{\percent} error bounds of \SI{2.33}{\meter}
horizontally (semi-major axis), \SI{3.37}{\meter} vertically, and
\SI{11.19}{\nano\second} in timing. The actual solution errors fall well within
these bounds. These results motivate further research in (i) a dedicated fused
LEO GNSS service as in~\cite{iannucci2022fusedLeo}, or (ii) a regional
third-party service deploying reference stations to relay corrections, as
in~\cite[Sec.~9.7.4]{rubinov2024leopnt}.
\section{Concept of Operations}
\label{sec:conops}

This section details the key assumptions underlying the fused LEO GNSS concept
regarding the CS, the SV constellation, and the UT. We then describe our
experimental setup, explaining how each component aligns with these assumptions.

\subsection{Fused LEO GNSS}

We assume an SV constellation primarily dedicated to broadband communications
but with a secondary mission to provide PNT services at minimal additional cost
to the user and minimal opportunity cost (e.g., reduced throughput) to the
communications network. Following~\cite{iannucci2022fusedLeo}, we refer to this
dual-purpose architecture as a \emph{fused LEO GNSS}. To distinguish this
architecture from medium Earth orbit systems (e.g., GPS, Galileo, BeiDou), we
refer to the latter simply as \emph{legacy GNSS}.

A fused LEO GNSS could rely on onboard legacy GNSS receivers for SV orbit
determination and clock disciplining. While an on-orbit receiver is not immune
to high-power terrestrial interference or severe space weather, leveraging
onboard GNSS offers a pragmatic near-term solution for existing constellations
such as Starlink and OneWeb, as it requires virtually no new deployment costs.
Moreover, receivers in LEO enjoy roughly \SI{30}{\decibel} of additional
protection from terrestrial GNSS interference compared with receivers at or
below \SI{10}{\kilo\meter} altitude.

In the long term, a robust fused LEO GNSS must rely on a network of ground
stations for precise orbit determination and clock monitoring. This network
(i.e., the CS) generates ephemeris and clock models and uplinks them to the SVs
for broadcast to end users, as with legacy GNSS.

The SVs themselves may operate with either a transparent or regenerative payload
architecture. To ensure sufficient multiple-access orthogonality, the service
provider retains full flexibility over the transmission scheme, including
waveform design, frequency channelization, beam directionality, and
polarization.

We assume the system operates as a LEO mega-constellation comprising thousands
of SVs (e.g., SpaceX's Starlink). This massive scale is critical, providing both
global communications coverage and the rich geometric diversity necessary for
robust PNT.

We model the UT as a consumer-grade device fully compatible with the network's
resource allocation scheme. If, for instance, the constellation transmits narrow
beams across multiple frequency channels, a fully fused UT (i.e., capable of all
the normal communications functions) must employ a phased-array antenna to
rapidly track and switch between these directional beams, and must be equipped
with the radio frequency (RF) chains required to process the scheduled channels.
The UT may be able to capture multiple frequency channels at once, increasing
its data rate and, through greater geometric diversity, improving its PNT
accuracy.

Crucially, demodulating symbols differs from decrypting user data. The CS and UT
may demodulate symbols transmitted by a given SV as much as necessary, but we
assume that payloads destined for other users remain encrypted and securely
inaccessible to the local UT.

A UT dedicated solely to PNT need only be capable of computing a PNT solution
from the network resources allocated for PNT services. This opens the
possibility for a distinct category of UT: the PNT-only device. Since the
signal-to-noise ratio (SNR) requirement for PNT is significantly lower than for
broadband communications, a service provider could feasibly offer a compact,
portable PNT-only UT with a low-gain antenna whose loss in gain relative to a
more directional antenna would be recovered through additional processing
gain. Reference~\cite{qin2026starlinktemplate} shows that Starlink, for example,
often sends packets that bear almost no information and could thus offer the
necessary processing gain and data to support PNT-only devices.

\begin{figure}[t]
  \centering
  \includegraphics[width=0.95\columnwidth]{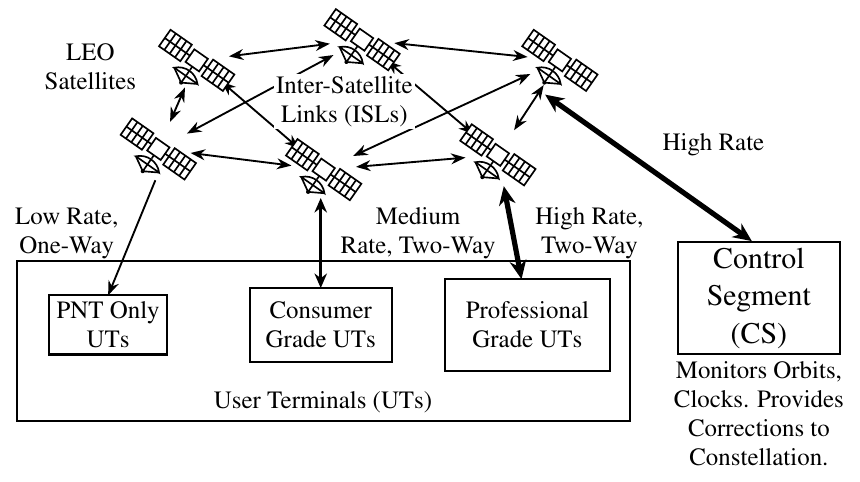}
  \caption{Fused LEO GNSS concept of operations with various user terminal
    types. Rate here refers to data rate required to send pilot sequences, PNT
    information, and other user-specific data.}
  \label{fig:fusedLEOConops}
\end{figure}

With this in mind, one could consider a taxonomy of various UT types: PNT-only
UTs, consumer-grade fused UTs capable of both PNT and broadband communications,
and professional-grade fused UTs capable of high-throughput broadband
communications and high-accuracy PNT. Each UT type would have different hardware
capabilities and SNR requirements. This taxonomy and how it fits in the network
as a whole are illustrated in Fig.~\ref{fig:fusedLEOConops}.

\subsection{Mock Implementation}
To prove the feasibility of the fused LEO GNSS concept, we demonstrate it
experimentally through a mock implementation with Starlink as the target
constellation.  This mock implementation retains all essential elements of the
fused approach.

In our implementation, the CS consists of a single reference station at a known
position, mounted on the roof of the parking garage adjoining the Aerospace
Engineering Building at the University of Texas at Austin. Equipped with a
high-gain antenna, the CS is capable of receiving full-bandwidth Ku-band
downlink signal data from a single Starlink SV at a time.  By processing the
digitized received signal, augmented with knowledge of its own position and
access to precise GPS time (GPST), the CS provides differential corrections to
the UT.

Our implementation's UT is a prototype PNT-only UT and is situated at a location
unknown to the CS.  Specifically, it comprises a Ku-band feedhorn antenna with a
half-power beamwidth that was empirically determined to be \SI{40}{\degree} at
\SI{10.35}{\giga\hertz}. Assuming a symmetric pencil beam at that beamwidth and
an antenna efficiency of 0.7, Kraus's approximation gives the antenna's gain to
be approximately \SI{12.6}{\dBi}~\cite[Eq.~2-9-4]{kraus88ant}. Our UT is
equipped with an RF front end capable of capturing only a \SI{25}{\mega\hertz}
real-time bandwidth---a small fraction of the total bandwidth of the Starlink
Ku-band downlink waveform. Our prototype UT is also limited to receiving data on
a single frequency channel at a time.

This configuration represents a conservative scenario, as the UT cannot reliably
and accurately measure TOAs based solely on existing Starlink pilot sequences
that support communications, such as the primary and secondary synchronization
sequences (PSS and SSS) revealed in~\cite{humphreys2023starlinkSignalStructure}.
Such a UT might be able to measure TOAs from a single band of the edge pilots
discovered in~\cite{qin2026starlinktemplate}, but these were unknown at the time
of our experiment. Alternatively, the CS could provide the detected so-called
T-codes as described in~\cite{qin2026starlinktemplate} to the UT, but these too
were unknown to the authors at the time of our experiment. Therefore, in our
experiment, the CS feeds the UT demodulated payload data that act as pilot data
to overcome weak SNR by additional processing gain; evaluating T-code-aided TOA
estimation and PNT is left to future work.  The UT also receives orbit and clock
corrections from the CS as if it were able to decode them from the downlink.

Because our demonstration's CS tracks and generates corrections for only a
single SV at a time, the UT cannot perform instantaneous single-epoch
multilateration. To establish observability, the UT must track Starlink SVs
sequentially, batching pseudorange measurements from each SV over its visibility
window. Over this time, the geometric diversity provided by a single SV's motion
can be equivalent to observing two or more SVs in the legacy GNSS regime.
Therefore, a PNT solution with fewer than four concurrent SVs is possible,
though the solution is non-instantaneous and is sensitive to UT clock drift.

\section{Signal Structure}
\label{sec:sys-concept-term}
This section summarizes relevant aspects of the Starlink frame structure, drawn
from~\cite{qin2025starlinktimingproperties},~\cite{humphreys2023starlinkSignalStructure},
and~\cite{qin2026starlinktemplate}. Additional concepts and terminology needed
to understand Starlink operation in general and Starlink-based PNT in particular
are also provided.

\subsection{Frame Structure}
\label{sec:frame-structure-1}

\begin{figure*}[t]
  \centering
  \includegraphics[width=0.75\textwidth]{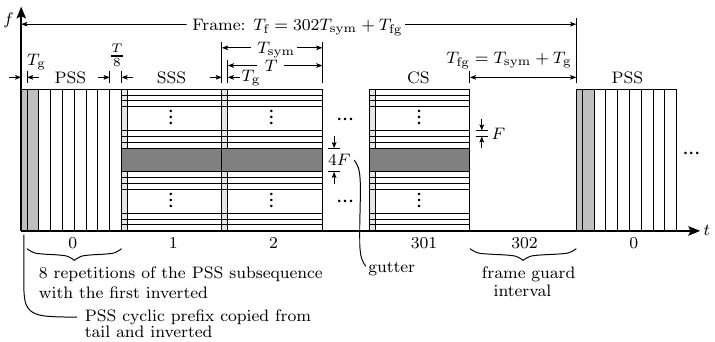}
  \caption{Frame layout for the Ku-band Starlink downlink along time-frequency
    dimensions, from~\cite{humphreys2023starlinkSignalStructure}.  Indices along
    the horizontal axis enumerate the 303 intervals that constitute a single
    frame.  The quantity $T$ is the useful (non-cyclic) OFDM symbol interval,
    $\Tg$ is the symbol guard interval (cyclic prefix), and $F$ is the
    subcarrier spacing.  Other quantities are defined in the text.}
  \label{fig:frame-diagram}
\end{figure*}

As shown in Fig.~\ref{fig:frame-diagram}, each Starlink downlink orthogonal
frequency division multiplexing (OFDM) frame consists of 302 intervals of length
$\Tsym = \SI{4.4}{\micro\second}$ plus a frame guard interval $\Tfg$, for a
total frame period of $\Tf = 1/\Ff$, where $\Ff = \SI{750}{\hertz}$ is the frame
rate.
Each frame begins with the PSS, which is natively represented in the time
domain, followed by the SSS, which is formatted as a standard quadrature
phase-shift keying OFDM
symbol. The final occupied symbol interval in each frame is the coda symbol
(labeled CS in Fig.~\ref{fig:frame-diagram}, not to be confused with the control
segment),
which is followed immediately by the frame guard interval. Immediately
subsequent frames may or may not be present,
depending on user demand.

The time-domain modulation sequence of the PSS and the frequency-domain symbol
sequence of the SSS were revealed
in~\cite{humphreys2023starlinkSignalStructure}. The 16 edge pilots (EPs), eight
subcarriers at each edge of the OFDM frame that are modulated with a
deterministic sequence of symbols, were presented
in~\cite{qin2026starlinktemplate}. The PSS, SSS, and EPs allow a Starlink
receiver to perform the channel estimation necessary to demodulate each OFDM
symbol in the frame. In particular, the PSS allows the receiver to precisely
identify the beginning of each frame, the SSS allows it to perform
equalization across all subcarriers, and the EPs allow estimation of
sample-clock-to-carrier divergence. The PSS and SSS have a fixed phase
relationship, which allows coherent integration across the combined PSS + SSS
interval, permitting accurate synchronization and carrier frequency offset
estimation, as shown in~\cite{qin2025starlinktimingproperties}.

\subsection{Assigned Beams}
At any given time, a user within Starlink's coverage area may have a direct line
of sight to hundreds of SVs above the horizon.
Each Starlink SV is capable of simultaneously directing multiple downlink beams
to terrestrial service cells,
the exact number depending on the SV version. A proprietary beam
assignment procedure assigns each beam to a $\sim$20-km-diameter service
cell~\cite{pekhterev2021bandwidth}.  A service cell may be illuminated by up to
128
beams simultaneously~\cite{fcc_gen2_upgrade_2026, fcc_sb25_157_factsheet}: eight
beams for each of the eight frequency channels
identified in~\cite{humphreys2023starlinkSignalStructure}, and two
polarizations.
We call the $\Nbi$ beams deliberately directed to the $i$th service cell
\emph{assigned beams}.

The number of distinct SVs casting assigned beams onto the $i$th service cell,
$\Nasi \leq \Nbi \leq 128$, may be less than $\Nbi$ because a given SV may
project multiple beams onto the same cell, each on a different channel or with a
different polarization (right-hand vs. left-hand circular polarization).  As
Starlink user density has increased over the past few years, $\Nasi$ and $\Nbi$
have generally increased.  The authors of~\cite{qin2023turret} showed that a
search
procedure prioritizing high-elevation SVs is effective at finding multiple
illuminating SVs.

\subsection{Fixed Assignment Interval}
Several publications have noted the existence of a 15-s Starlink network
reconfiguration interval~\cite{qin2023turret, huston2024transportProtocol,
  mohan2024multifaceted, grayver2024position}. Because all primary beam
assignments remain fixed over the duration of this interval, we call it the
\emph{fixed assignment interval} (FAI). Within an FAI, a beam may nonetheless
briefly serve, or ``moonlight'' on, service cells other than its primary
assignment. FAI boundaries appear to be approximately aligned with
GPST~\cite{qin2023turret}. As discussed
in~\cite{qin2025starlinktimingproperties}, noticeable transitions in Starlink
frame timing behavior occur at FAI boundaries.

\section{Clock Models}
\label{sec:clock-models}

This section introduces the Starlink clock models described
in~\cite{qin2025starlinktimingproperties} and~\cite{qin2026starlinktemplate}.
These models are necessary to derive
the clock corrections provided by the CS in Section~\ref{sec:ephClkCorr}.
\subsection{Beam-Specific Clock Cascade Model}
\label{sec:beam-specific-clock}
\begin{figure}[t]
  \centering
  \includegraphics[width=\columnwidth]{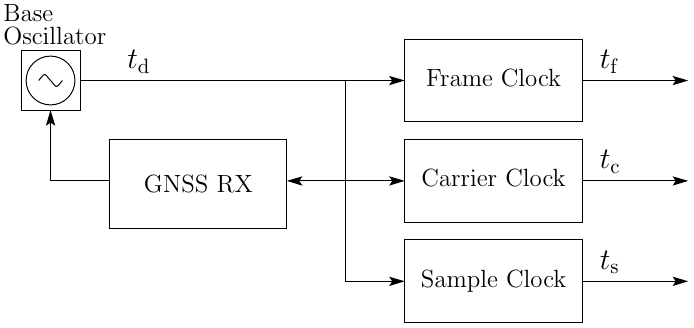}
  \caption{Cascading clock model for the Starlink transmitter,
  from~\cite{qin2026starlinktemplate}.}
  \label{fig:clock-model}
\end{figure}

Whereas a single clock drives both the information symbols and the carrier of a
direct-sequence spread-spectrum waveform, separate clocks drive those of the
Starlink downlink.
The cascading clock model, shown schematically in Fig.~\ref{fig:clock-model},
explains the code-carrier divergence observed in the received signal.
A GNSS-disciplined base time $\td$ drives the downstream frame, sample, and
carrier clocks, which independently drive signal generation onboard the Starlink
SV.
The frame clock sets the transmission start time of each frame, the sample clock
drives the digital-to-analog converter that reconstructs the digital waveform as
an analog baseband waveform, and the carrier clock drives generation of the
carrier signal that mixes the baseband OFDM waveform up to RF.

Time according to these three clocks is related to true time $t$, or time
according to an ideal clock, by
\begin{align}
  t & = \tf(t) - \dtf(t) \\
  \label{eq:1d-tc}
  t & = \tc(t) - \dtc(t) \\
  \label{eq:1d-ts}
  t & = \ts(t) - \dts(t)
\end{align}
where $\dtf(t)$, $\dts(t)$, and $\dtc(t)$ are the frame, sample, and carrier
clock offsets, respectively. In this paper, true time is assumed to be realized
by GPST.

The time derivatives of $\dtf(t)$ and of $\dtc(t)$ are called the
frame and carrier clock drift. The time derivative of $\dts(t)$ is called the
sample clock drift.
The sample clock offset and drift impart errors on estimates of $\dtf(t)$, but
these errors are identical at the CS and UT. Hence, this paper treats the sample
clock offset and drift as components of $\dtf(t)$ and does not discuss them
further.

\begin{figure}[t]
  \centering
  \includegraphics[width=\columnwidth]
  {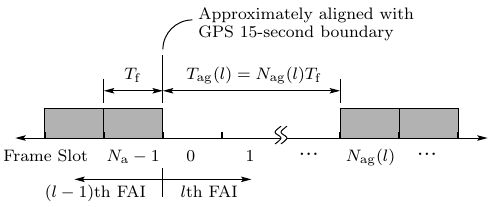}
  \caption{Frame sequence timing diagram showing the transition from the
  $(l-1)$th FAI
    to the $l$th FAI.}
  \label{fig:frameTimeDiagram}
\end{figure}

\subsection{Discrete-Time Frame Clock}
The frame sequence timing diagram in Fig.~\ref{fig:frameTimeDiagram} offers
further details about the frame clock.  Each frame as transmitted has a duration
of exactly $\Tf$ according to the frame clock.  Within each FAI, the frame slot
index increments from $0$ to $\Na - 1$, with $\Na = 11250$ being the number of
frame slots in one FAI.  Each FAI starts at the beginning of frame slot 0 and
lasts $\Na \Tf =$ \SI{15}{\second}.  The interval of unoccupied frame slots at
the beginning
of the $l$th FAI, called the FAI guard interval $\Tag(l) = \Nag(l) \Tf$, spans a
variable number of frame slots $\Nag(l)$.  Note that, for any FAI index $l$,
frame slot $\Nag(l)$ is occupied by definition, but other frame slots may not be
occupied.

Let $\tf(i,l,m)$ be the frame clock time for the $i$th Starlink SV at the
instant when the frame in the $m$th frame slot of the $l$th FAI begins to pass
through the phase center of the SV's downlink antenna, where $l$ and $m$ are
zero-based indices. By definition, we take this to be
\begin{align}
  \label{eq:tflm-def}
  \tf(i,l,m) \define 15l +  m\Tf
\end{align}
Let $t^*(i,l,m)$ and $\dtf(i,l,m)$ be the corresponding true time and frame
clock offset.  Then
\begin{align}
  \label{eq:tflm-def-true}
  t^*(i,l,m) = \tf(i,l,m) - \dtf(i,l,m)
\end{align}

\subsection{Receiver Clock}
Now consider the clock of a receiver (e.g., the UT) tracking signals from
the Starlink downlink.  The receiver clock time $\tr$ is related to true time
$t$ by
\begin{equation}
  \label{eq:receiverClock}
  t = \tr(t) - \delta \tr(\tr)
\end{equation}
The receiver clock offset $\delta\tr(\tr)$ is represented as a function of $\tr$
because it is natively ordered in receiver time in the course of solving for a
position and time solution.  The time derivative of $\delta \tr$ with respect to
$t$, denoted $\delta\dot{t}_\text{r}(\tr(t))$, is called the receiver clock
drift.

Let $\tr(i,l,m)$ be the time of reception, according to the receiver clock, of
the frame transmitted by the $i$th SV at true time $t^*(i,l,m)$.  Let
$\delta\tr(i,l,m)$ be the corresponding receiver clock offset and $t_*(i,l,m)$
be the corresponding true time of reception.  More precisely, $\tr(i,l,m)$ is
the receiver clock time at which the frame transmitted at true time $t^*(i,l,m)$
from the $i$th SV's downlink antenna's phase center first reached the receiver
antenna's phase center.  The receipt time $\tr(i,l,m)$ can be related to
$t_*(i,l,m), t^*(i,l,m)$, and $\tf(i,l,m)$ by
\begin{align}
  t_*(i,l,m) & = \tr(i,l,m) - \delta\tr(i,l,m)                               \label{eq:tgt-rx}   \\
  t^*(i,l,m) & = \tr(i,l,m) - \delta\tr(i,l,m) - \dTOF(i,l,m)                  \label{eq:tgt-tx} \\
  \tf(i,l,m) & = \tr(i,l,m) - \delta\tr(i,l,m) - \dTOF(i,l,m) \nonumber                          \\
             & \quad + \dtf(i,l,m) \label{eq:tflm-comp}
\end{align}
where $\dTOF(i,l,m)$ is the frame's time of flight from transmission to
reception,
expressed as an interval in true time.

\section{Data Aiding}
\label{sec:dataAiding}
The inclusion of additional low-entropy elements beyond the PSS, SSS, and EPs in
the local replica generated by the UT improves both detection and TOA
measurement performance. Hereafter, the EPs are ignored because the frequency
alignment of our UT's \SI{25}{\mega\hertz} band did not capture them. This
section discusses four data-aiding strategies that affect the UT's processing
gain and TOA estimation error: (1) no data aiding, in which the UT correlates
against the PSS and SSS alone (PSS + SSS); (2) aiding with subcarrier pilots
allocated optimally according to the ZZB across a range of received SNRs, in
addition to the PSS and SSS (Optimal Pilots); (3) aiding with subcarrier pilots
allocated optimally according to the ZZB at a single worst-case received SNR, in
addition to the PSS and SSS (Worst-Case Allocations); and (4) aiding with the
entire downlink frame within the UT's front-end bandwidth, either decoded by the
UT or provided by the CS (Full Frame).
\subsection{TOA Estimation Error Bounds}
In high-SNR regimes, the Cram\'er--Rao bound (CRB) is a tight bound on TOA
estimation error. The CRB on TOA error variance for a signal with Fourier
transform $S(f)$, energy $E$, and noise spectral density $N_0$
is~\cite[Eq.~(11.15)]{SkolnikRadarSystems}
\begin{align}
  \sigma_{\tau}^2 \geq \frac{1}{2\gamma^2 \sfrac{E}{N_0}}
\end{align}
where $\gamma^2$ is the effective squared bandwidth defined as
\begin{align}
  \gamma^2 \define \frac{\int_{-\infty}^{\infty}(2\pi f)^2 |S(f)|^2 \,df}{\int_{-\infty}^{\infty} |S(f)|^2 \,df}
\end{align}
However, in low-SNR regimes, the CRB fails to capture the thresholding effects
caused by sidelobes, whereas the ZZB captures them. Derivations of the ZZB for
OFDM TOA estimation are given
in~\cite{graff2024purposeful,graff2024ziv,graff2025data}.
\subsection{PSS + SSS}
If the UT can detect the Starlink signal with a local replica of the PSS + SSS
alone, TOA measurements obtained this way should provide reasonable accuracy.
In~\cite{qin2025starlinktimingproperties}, the authors show that sub-centimeter
accuracy is theoretically achievable at moderate SNR with the full
\SI{240}{\mega\hertz} channel bandwidth. Because our UT is limited to a
\SI{25}{\mega\hertz} bandwidth, a new calculation is necessary.
The PSS signal given in~\cite[Eq.~(37)]{humphreys2023starlinkSignalStructure}
has a Fourier transform
\begin{align}
  S_{\text{PSS}}(f) \define \frac{1}{\Fs}\sum_{k=-\Ng}^{\N-1} \text{rect}\left(\sfrac{f}{\Fs}\right)\exp{\left(-j2\pi (k+\Ng) \sfrac{f}{\Fs}\right)} p_k
\end{align}
and, when sampled at \SI{25}{\mega\hertz}, an effective bandwidth of
$\gamma_{\text{PSS}} = \SI{3.9977e7}{\radian\per\second}$.\footnote{Effective
bandwidth is reported in units of \SI{}{\hertz}
in~\cite{qin2025starlinktimingproperties}, but should have been reported in
\SI{}{\radian\per\second}. The TOA measurement bounds themselves are
unaffected.}
Similarly, the SSS signal given
in~\cite[Eq.~(40)]{humphreys2023starlinkSignalStructure} has a Fourier transform
\begin{align}
  S_{\text{SSS}}(f) & \define \sum_{k=-\sfrac{\N}{2}}^{\sfrac{\N}{2}-1} X_{m1k}\exp{\left(-j2\pi \F \Tg k\right)}  G_{\text{s}}(f-\F k) \\
  G_{\text{s}}(f)   & \define \Tsym \sinc\left(\Tsym f\right)\exp\left(-j\pi\Tsym f\right)
\end{align}
and, when sampled at \SI{25}{\mega\hertz}, an effective bandwidth of
$\gamma_{\text{SSS}} = \SI{4.6063e7}{\radian\per\second}$.
The combined PSS + SSS signal in a single frame has the Fourier
transform $S_{1}(f) = S_{\text{PSS}}(f) + \exp\left(-j2\pi \Tsym f\right)
  S_{\text{SSS}}(f)$ and, when sampled at \SI{25}{\mega\hertz}, an effective
  bandwidth of $\gamma_{\text{PSS+SSS}} = \SI{4.3283e7}{\radian\per\second}$.
  Importantly, because the PSS and SSS are already transmitted in every frame, a
  UT that can form TOA measurements from the PSS + SSS alone imposes no loss of
  data capacity on the network.

\subsection{Optimal Pilots}
To compensate for possibly low received SNR at the UT (e.g., for a PNT-only UT
or under interference), the CS may transmit demodulated data from the Starlink
frame to the UT. The UT can then include these known resources in addition to
the PSS and SSS in its local signal replica, allowing a more robust and accurate
TOA estimate than would be achievable with the PSS + SSS alone. In our
experiment, this technique constitutes a form of data aiding. In practice,
however, the Starlink network would designate these data resources as known
pilots, and the UT would have \textit{a priori} knowledge of them.

In this scheme, it is critical that the overhead required to transmit these data
to the UT be minimized, ensuring reliable and low-latency data transmission. To
achieve this, the UT receives only a small selection of demodulated data
resources corresponding to a sparse subset of subcarriers that span the UT's
available bandwidth. However, the specific selection of subcarriers affects
the TOA autocorrelation function. If the UT is provided with a suboptimal
allocation, its autocorrelation function may exhibit ambiguities or high
sidelobes, leading to misdetections and increased TOA estimation
errors. Conversely, if the UT is provided with an optimal allocation, the
autocorrelation function will exhibit a sharp mainlobe and adequately suppressed
sidelobes, improving the TOA accuracy for the given SNR.

\begin{figure}[t]
  \centering
  \includegraphics[width=\columnwidth]{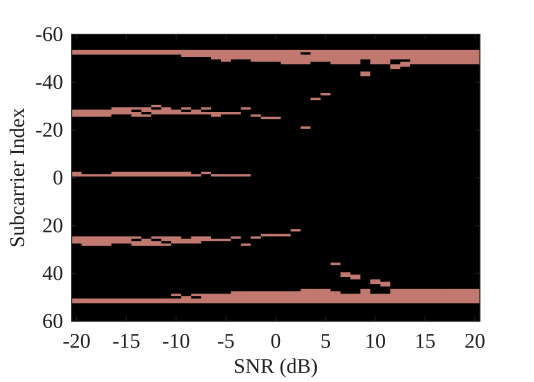}
  \caption{Optimized selections of 12 out of 106 available subcarriers for data
  aiding as a function of pre-correlation SNR.}
  \label{fig:subcarrier_allocations}
\end{figure}

We generated an optimized set of subcarrier selections for pre-correlation SNRs
ranging from \qtyrange{-20}{20}{\decibel} with the noncoherent
integer-constrained ZZB-minimization techniques of~\cite{graff2024ziv}. In this
minimization, the ZZB assumed an \textit{a priori} TOA window with a duration
equal to that of the cyclic prefix. The returned allocations are near-optimal,
achieving a ZZB TOA variance within \SI{5}{\percent} of that of the true optimal
selection. Each selection consists of 12 out of 106 available subcarriers within
the UT's \SI{25}{\mega\hertz} bandwidth. The choice of 12 subcarriers is
somewhat arbitrary; other choices trade additional data-aiding overhead for
improved TOA accuracy. These subcarrier selections are shown in
Fig.~\ref{fig:subcarrier_allocations}. At low SNR, the selection is more evenly
distributed across the available bandwidth, while at high SNR the selection
favors the extremities of the band, thereby maximizing the effective squared
bandwidth.

The CS can select subcarriers based on the UT's SNR, demodulate all data
resources in the Starlink frame corresponding to these subcarriers, and transmit
the decoded data to the UT as though they were pilots sent by the network. If
the Starlink network were to implement this, the transmitting SV would need to
sacrifice 12 subcarriers per OFDM symbol in each frame. Given that each Starlink
OFDM symbol consists of 1024 subcarriers, the absolute data-rate loss for the
entire frame would be \SI{1.17}{\percent}. In an operational network serving
many UTs, each at a different SNR, each UT would report signal-quality telemetry
to the SV, which would then allocate the pilots optimized for that UT's SNR to
the appropriate subcarriers within that UT's own resource allocation.

For the CRB on TOA measurement errors, the Fourier transform of the entire frame
at a given pre-correlation SNR is $S_{\text{O}}(f) = S_1(f) +
\sum_{k=2}^{300}\exp\left(-j2\pi k \Tsym f\right) S_{\text{S}}(f)$, where
$S_{\text{S}}(f)$ represents the optimal subcarrier allocation for a given
symbol at a specific SNR. Assuming that the SNR remains constant over the course
of a frame, we can compute $S_{\text{O}}$ for each SNR, determine the effective
bandwidth, and calculate the bound on TOA accuracy.

These allocations are optimal on a per-symbol basis. Consequently, a TOA
measurement computed with a single optimal subcarrier allocation across the
entire frame may not be optimal. For the purposes of this paper, we assume that
the difference is minor.

\subsection{Worst-Case Allocations}
Because the optimal subcarrier allocations require the network to know each UT's
SNR, this data-aiding policy could prove operationally burdensome.
Alternatively, the network could always allocate as pilots the subcarriers
optimized for the lowest SNR. In this way, all UTs are protected in the
worst-case SNR regime.
However, the performance loss that this allocation would incur in the best-case
SNR regime must also be quantified. We quantify it as for the optimal pilots,
but with the allocation optimized for the worst-case pre-correlation SNR of
\SI{-20}{\decibel} held fixed at every SNR. Holding this allocation fixed
increases the TOA root mean squared error (RMSE) CRB and ZZB by at most
\SI{40}{\percent}
relative to the SNR-specific allocations. This increase is most pronounced
relative to the allocations optimized for SNRs above \SI{-10}{\decibel}, but the
absolute increase in TOA RMSE remains below \SI{3}{\centi\meter}.
Therefore, depending on the UTs' accuracy requirements, a single worst-case
subcarrier allocation could serve all UTs at minimal cost to TOA accuracy.

\subsection{Full Frame}
Finally, the UT may form a TOA measurement from the entire frame within its
bandwidth. This could happen in two ways. In the first, the network provider may
choose to dedicate the entire bandwidth of the UT to pilot subcarriers. The UT
would then be designated as PNT-only, as no data would remain to decode, and it
would have to obtain certain PNT data from a side channel or from an earlier
frame modified for low SNRs. In this case, the SV would allocate 106 subcarriers
per symbol per frame, assuming a UT with a \SI{25}{\mega\hertz} bandwidth. The
absolute data-rate loss incurred by this allocation would be
\SI{10.35}{\percent}.

Alternatively, if the UT were not PNT-only but instead had sufficient gain to
permit decoding data, it could form a local replica from the decoded data and
measure TOA as described in~\cite{graff2025data}. This
\textit{decision-directed} scheme would incur no cost to the network but would
require a more complex UT. Additionally, any decoding errors made by the UT
would affect TOA measurement accuracy.

To compute the CRB on TOA measurement errors, we assume that the entire frame
within the UT's front-end bandwidth is transmitted as pilots. The Fourier
transform of the entire frame is $S_{\text{F}}(f) = S_1(f) +
\frac{1}{\Fs}\sum_{k=2}^{300}\exp\left(-j2\pi k \Tsym f\right)
\text{rect}\left(\sfrac{f}{\Fs}\right)$, assuming that the Fourier transform of
the remaining symbols in the frame is rectangular. The effective bandwidth of
the entire frame under this assumption, when sampled at \SI{25}{\mega\hertz}, is
$\gamma_{\text{F}} = \SI{4.6434e7}{\radian\per\second}$.

\begin{figure}[t]
  \centering
  \includegraphics[width=\columnwidth]{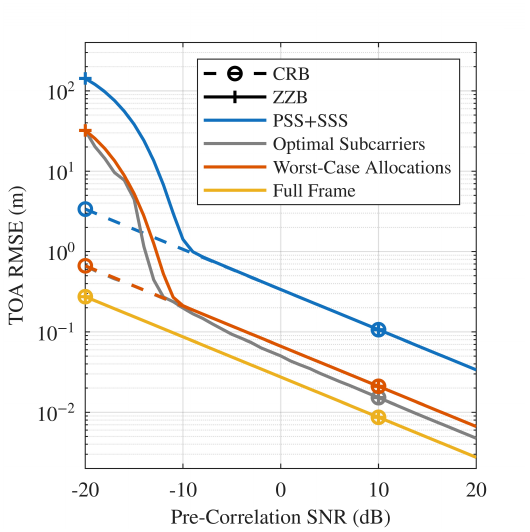}
  \caption{TOA RMSE CRBs and ZZBs for four levels of data aiding to a UT with a
  front-end bandwidth of \SI{25}{\mega\hertz}. Relying on the PSS + SSS alone
  incurs no loss of data capacity, the 12-subcarrier optimal and worst-case
  pilot allocations each incur a \SI{1.17}{\percent} loss, and aiding with the
  entire \SI{25}{\mega\hertz} band incurs a \SI{10.35}{\percent} loss. Only the
  optimal allocations require knowledge of the UT's SNR. Thresholding effects
  begin at \SI{-9}{\decibel}, \SI{-11}{\decibel}, and \SI{-12}{\decibel} for the
  PSS + SSS, worst-case, and optimal pilot allocations, respectively, whereas
  the full frame exhibits no thresholding above \SI{-20}{\decibel}. The optimal
  pilot allocations thus provide \SI{3}{\decibel} of thresholding protection
  relative to the PSS + SSS and \SI{1}{\decibel} relative to the worst-case
  allocations.}
  \label{fig:toarmsebounds}
\end{figure}

Fig.~\ref{fig:toarmsebounds} plots RMSE CRBs and ZZBs for the four strategies
discussed in this section over the range of \qtyrange{-20}{20}{\decibel}. As
expected, the PSS + SSS provide the worst TOA accuracy but still achieve
sub-meter accuracy at relatively low SNR. The optimal and worst-case pilot
allocations yield significant performance improvements, both theoretically
reaching sub-centimeter accuracy at high SNR. The full frame offers the best
accuracy, theoretically reaching sub-centimeter accuracy at an SNR approximately
\SI{5}{\decibel} lower than the optimal pilot allocations and \SI{8}{\decibel}
lower than the worst-case allocations.

\subsection{Data Aiding Criteria}
The appropriate data-aiding strategy depends on the UT's application
requirements, its hardware, and its current signal environment. Because data
aiding consumes network resources, the decision rests with the CS, which must
weigh each UT's needs against the capacity reserved for communications. The UT's
role is to inform this decision.

To that end, each UT reports signal-quality telemetry, such as its estimated
pre-correlation SNR, to the network. Because this estimate is derived from the
received signal rather than from nominal hardware parameters, it captures SNR
degradation due to jamming or other interference as it occurs. Before its first
detection, a UT has no such estimate, and the CS may instead assume a nominal
SNR based on the UT's hardware class (e.g., front-end bandwidth and antenna
gain). In either case, the CS formulates detection as a statistical hypothesis
test~\cite{komodromos2023IONsimulator} to calculate the minimum processing gain
necessary to satisfy target probabilities of detection and false alarm at the
UT's SNR. It then selects the data-aiding strategy that safely exceeds this
detection threshold while minimizing communications overhead, and revises its
selection as the reported SNR changes.

For UTs demanding high-precision TOA measurements, the CS may allocate optimally
spaced pilots or the full frame, trading data capacity for navigation accuracy.
Conversely, for a UT that prioritizes communications throughput and tolerates
lower TOA precision, the CS may rely solely on the PSS + SSS, provided these
supply sufficient processing gain, at no additional network overhead.

\section{Satellite Clock and Ephemeris Corrections}
\label{sec:ephClkCorr}
This section introduces the pseudorange measurement model and identifies where
errors due to the ephemeris and transmitter clock may arise. Additionally, we
explore how these errors are mitigated in practice and in our specific
experiment.

\subsection{Pseudorange Measurement Model}
In a fused LEO GNSS based on Starlink, a UT would apply (\ref{eq:tflm-def}) to
obtain the time of transmission (TOT), according to the $i$th SV's frame clock,
of the $m$th frame in the $l$th FAI.  For its part, the network would send the
frame clock offset $\dtf(i,l,m)$ so that the UT could obtain the corresponding
TOT in GPST, as in (\ref{eq:tflm-def-true}).  Likewise, in our experiment, the
CS estimates the
frame clock offset $\dtf(i,l,m)$ and provides it to the UT, as described below.

Rearranging (\ref{eq:tflm-comp}) to isolate $\dtf(i,l,m)$ yields
\begin{equation}
  \label{eq:isolated_df}
  \dtf(i,l,m) = \tf(i,l,m) - \tr(i,l,m) + \delta\tr(i,l,m) + \dTOF(i,l,m)
\end{equation}
If the CS can calculate, measure, or model all the quantities on the right-hand
side of this equation, it obtains an estimate of $\dtf(i,l,m)$ that it can
pass to the UT.  To obtain $\tf(i,l,m)$, it calculates nominal frame timing
according to (\ref{eq:tflm-def}).  This calculation is unambiguous because, as
explained in~\cite{qin2025starlinktimingproperties}, the
CS can mark the end of an FAI by observation and count frame slots within the
subsequent FAI.

The CS measures $\tr(i,l,m)$.  Let $\Mil$ be the set of
occupied-frame indices for the $l$th FAI of the $i$th SV.  The CS extracts a
sequence of
highly precise frame TOA measurements corresponding to $\Mil$ with the
maximum-likelihood (ML) methods described in~\cite{qin2025toaDopplerPlans}.  For
$m \in \Mil$, these are modeled as
\begin{align}\label{eq:cstoa}
  \ttr(i, l, m) & = \tr(i,l,m) + \wwr(i,l,m)
\end{align}
where $\wwr(i,l,m)$ is zero-mean Gaussian measurement error with variance
$\sigma_\text{r}^2(i,l,m)$.  As explained in~\cite{qin2025toaDopplerPlans}, if
the CS has access to the full \SI{240}{\mega\hertz} channel bandwidth and an
antenna with
sufficient gain to provide a pre-correlation SNR of \SI{14}{\decibel} or more,
then ML TOA
estimation over a full frame is remarkably precise, with range-equivalent RMSE
$c \sigma_\text{r}(i,l,m)$ less than one centimeter, where $c$ is the speed of
light in a vacuum.

To estimate $\delta\tr(i,l,m)$, the CS employs simultaneous Starlink and GNSS
signal capture, accounting for all cable delays, as described
in~\cite{qin2025starlinktimingproperties}.  Finally, the CS models
$\dTOF(i,l,m)$
from SV ephemerides, as also described
in~\cite{qin2025starlinktimingproperties}.

Unlike in~\cite{qin2025starlinktimingproperties}, meter-accurate ephemerides
based on observables from GNSS receivers onboard the Starlink SVs were not
available for this paper's demonstration.  Instead, we relied on publicly
accessible Starlink ephemerides from
\href{https://www.space-track.org/auth/login}{space-track.org},
which are accurate (at least periodically) to within about
\SI{10}{\meter}~\cite{liu2024maneuver}. These ephemeris errors introduced small
errors into our modeled $\dTOF(i,l,m)$.
Additionally, the estimated $\delta\tr(i,l,m)$ was accurate only to the
sub-nanosecond level.
The CS also replaces $\tr(i,l,m)$ in (\ref{eq:isolated_df}) with its measurement
$\ttr(i, l, m)$.
Thus, the estimated $\dtf(i,l,m)$ sent from the CS to the UT, denoted
$\dtfb(i,l,m)$, is modeled as
\begin{align}
  \label{eq:df_estimated_model}
  \dtfb(i,l,m) & = \dtf(i,l,m) + e_\text{r}(i,l,m)/c                                                               \\ \nonumber
               & \quad - \delta t_{\text{atm},\text{r}}(i,l,m) + \epsilon_{\delta t_\text{r}}(i,l,m) - \wwr(i,l,m)
\end{align}
where $e_\text{r}(i,l,m)$, given in meters, is the range error (predicted minus
true) due to ephemeris errors at the CS, $\delta t_{\text{atm},\text{r}}(i,l,m)$
describes atmospheric delays along the line-of-sight path at the CS, and
$\epsilon_{\delta t_\text{r}}(i,l,m)$ is the CS clock estimation error.

We now turn our attention to the UT, which calculates TOA just as the CS does
but with a possibly reduced bandwidth and, in the case of a PNT-only UT, with ML
estimation based on additional CS-provided pilot data, which provides the ML
estimator a boost in processing gain to compensate for the UT's low-gain
antenna.  Let $\ttu(i,l,m)$ denote the UT's measured frame TOA for the
corresponding $i$, $l$, and $m \in \Mil$, which is modeled as
\begin{align}\label{eq:cttoa}
  \ttu(i, l, m) & = \tu(i,l,m) + \wwu(i,l,m)
\end{align}
where $\tu(i,l,m)$ is the true TOA according to the UT's clock, and
$\wwu(i,l,m)$ is zero-mean Gaussian measurement error with variance
$\sigma_\text{u}^2(i,l,m) \geq \sigma_\text{r}^2(i,l,m)$.  The UT's pseudorange
measurement, defined as
\begin{equation}
  \label{eq:prDefinition}
  \rho(i,l,m) = c \left[\ttu(i,l,m) - \tf(i, l, m)\right]
\end{equation}
may be modeled as
\begin{align}\label{eq:prModel}
  \rho(i,l,m) & = r_\text{u}(i,l,m) + c \delta\tu(i,l,m) - c \dtf(i,l,m)        \\ \nonumber
              & \quad + c \delta t_{\text{atm},\text{u}}(i,l,m) + c \wwu(i,l,m)
\end{align}
where $r_\text{u}(i,l,m)$ is the true range between the $i$th Starlink SV at the
time of transmission
and the UT at the time of reception, $\delta\tu(i,l,m)$ is the UT clock error,
and $\delta t_{\text{atm},\text{u}}(i,l,m)$
denotes atmospheric delays along the line-of-sight path at the UT.

The UT measures $\rho(i,l,m)$ from a nominal $\tf(i,l,m)$ based on the frame
rate and a coarse initial $\tf(0,0,0)$.
Then, to estimate its state, the UT first predicts $\rho(i,l,m)$ using its
knowledge of its state and $\dtfb(i,l,m)$
as provided by the CS. Let $\bar{\rho}(i,l,m)$ be the estimator's predicted
pseudorange, which is formed from the same
ephemerides whose errors cause $e_\text{r}(i,l,m)$ to appear in
(\ref{eq:df_estimated_model}).
For simplicity, we assume the UT has perfect knowledge of its position and
clock.
The predicted pseudorange is then
\begin{equation}
  \label{eq:rhoPred}
  \bar{\rho}(i,l,m) = r_\text{u}(i,l,m) + e_\text{u}(i,l,m) + c \delta\tu(i,l,m) - c \dtfb(i,l,m)
\end{equation}
where $e_\text{u}(i,l,m)$ is the range error (predicted minus true) due to
ephemeris errors at the UT.
The pseudorange residual $\delta\rho(i,l,m) \define \rho(i,l,m) -
\bar{\rho}(i,l,m)$ is then
\begin{align}
  \label{eq:psrResidual}
  \delta\rho(i,l,m) & = \eta(i,l,m) + c \delta t_\text{atm}(i,l,m) + c \epsilon_{\delta t_\text{r}}(i,l,m) \\
                    & \quad + w_\rho(i,l,m) \nonumber
\end{align}
where $\eta(i,l,m) \define e_\text{r}(i,l,m) - e_\text{u}(i,l,m)$ is the
differential ephemeris error, which satisfies
$|\eta(i,l,m)| \ll |e_\text{u}(i,l,m)| \approx |e_\text{r}(i,l,m)|$ for short
baselines, $\delta t_\text{atm}(i,l,m) \define \delta
t_{\text{atm},\text{u}}(i,l,m) - \delta t_{\text{atm},\text{r}}(i,l,m)$
is the differential atmospheric error, and $w_\rho(i,l,m) \define
c\left[\wwu(i,l,m) - \wwr(i,l,m)\right]$ is zero-mean Gaussian
measurement error with variance $c^2 \left[ \sigma_\text{r}^2(i,l,m) +
\sigma_\text{u}^2(i,l,m) \right]$.

The terms $\eta(i,l,m)$, $\delta t_\text{atm}(i,l,m)$, and $\epsilon_{\delta
t_\text{r}}(i,l,m)$ are nearly constant over an FAI. Dropping the frame index
$m$, we therefore define the joint error term
\begin{equation}
  \zeta(i,l) \define \eta(i,l) + c \delta t_\text{atm}(i,l) + c \epsilon_{\delta t_\text{r}}(i,l)
\end{equation}
Finally, we rewrite (\ref{eq:psrResidual}) as
\begin{equation}
  \label{eq:finalRes}
  \delta\rho(i,l,m) = \zeta(i,l) + w_\rho(i,l,m)
\end{equation}
\section{Estimation}
\label{sec:estimation}
Section~\ref{sec:ephClkCorr} developed the pseudorange residual model and
identified its error terms.  This section develops the estimator that
recovers the UT position and clock trajectory from a collection of such
residuals.  It presents
the estimator formulation, the empirical measurement noise covariance, the
oscillator process model that couples measurements separated in time, and the
batch solution and its covariance.

An operational UT would typically form an instantaneous PNT solution from
simultaneous
residuals, for all of which the receiver clock error acts as a common bias.  Our
experimental data differ in two respects: they span several
minutes, and each SV appears in a single brief pass separated from the others by
long gaps.  Across such gaps the UT oscillator may wander more than the
measurement noise, so neither a single bias nor a bias-and-drift pair describes
it faithfully.  We therefore retain a clock bias and a clock drift state at
every
measurement epoch and couple consecutive epochs through the oscillator's
stochastic model.  The result is a weighted nonlinear batch least squares
(WNBLS) problem over the UT position and the full clock trajectory.

\subsection{Estimator Formulation}

In what follows, we process a single FAI per SV, allowing us to drop
the FAI index $l$ such that $\mathcal{M}_{il} \define \mathcal{M}_i$.  Let $N_z
=
  \sum_{i=0}^{n-1} |\mathcal{M}_i|$ denote the total number of aggregated
pseudoranges.

We then aggregate the $N_z$ pseudoranges from $n$ SVs into a single observation
vector
$\vb{z}$
\begin{equation}
  \label{eq:obsVec}
  \vb{z} = \begin{bmatrix}
    \vb{\rho}\T(0) & \vb{\rho}\T(1) & \dots & \vb{\rho}\T(n-1)
  \end{bmatrix}\T
\end{equation}
where $\vb{\rho}(i) = [\rho(i, 0), \dots, \rho(i, |\mathcal{M}_i| - 1)]\T$
contains the $|\mathcal{M}_i|$ sequential frame pseudoranges for the
$i$th SV. Similarly, we construct the predicted pseudorange vector
$\bar{\vb{z}}$
\begin{equation}
  \label{eq:predVec}
  \bar{\vb{z}} = \begin{bmatrix}
    \bar{\vb{\rho}}\T(0) & \bar{\vb{\rho}}\T(1) & \dots & \bar{\vb{\rho}}\T(n-1)
  \end{bmatrix}\T
\end{equation}
where $\bar{\vb{\rho}}(i) = [\bar{\rho}(i,0), \dots,
\bar{\rho}(i,|\mathcal{M}_i| - 1)]\T$.
The estimator operates on the residual vector $\delta \vb{z} = \vb{z} -
\bar{\vb{z}}$ which is stacked as in
(\ref{eq:obsVec}) and (\ref{eq:predVec}), but with $\delta \vb{\rho}(i)=[\delta
\rho(i,0),\dots, \delta \rho(i,|\mathcal{M}_i| - 1)]\T$.
At the true position and clock offset $\delta t_\text{u}$, the pseudorange
residual of the $m$th frame of the $i$th SV is
$\delta \rho(i,m) = \zeta(i) + w_\rho(i,m)$ as in (\ref{eq:finalRes}).

The residual vector exposes the joint error term $\zeta(i)$ to the estimator.
This term warrants attention because it determines whether
per-SV error states are required.  In our prior work~\cite{morgan2025mockplans},
the differential ephemeris error $\eta(i)$ dominated
$\zeta(i)$ and varied from SV to SV in a largely unmodelable way.
That work drew on TLE ephemerides, which at times carry kilometers of error and
therefore leave tens of meters of differential error, so $\eta(i)$ had to enter
the state vector.  The present work draws instead on SpaceX's own public
ephemerides, whose errors are smaller, and the
short CS-to-UT baseline cancels most of what remains.  Applying the differential
ephemeris error estimation techniques
of~\cite{morgan2025mockplans,komodromos2025NetworkPlans} with the covariances
published alongside the ephemerides reveals that $\eta(i)$ is at the
few-centimeter level.  The differential atmospheric term is likewise small:
ionospheric delay is slight at Ku band, and the tropospheric delays at the two
receivers remain heavily correlated across a short baseline, leaving errors of a
few centimeters~\cite{lawrence2006tropodecorrelation}.  The CS clock estimation
error $\epsilon_{\delta t_\text{r}}(i)$ therefore dominates $\zeta(i)$; the GNSS
receiver at the CS reported this error to have a $1\sigma$ bound of
$\sigma_{\zeta} = \SI{15}{\centi\meter}$.
Hence $\zeta(i)$ sits at the level of a few tens of centimeters.

Two features of $\zeta(i)$ permit its absorption into the clock states.  First,
it changes only when a new SV appears.  Second, and because of the circumstances
of our experiment, a new SV appears only after a long temporal gap, across which
an OCXO accumulates timing error of the same order.  A step in $\zeta(i)$ is
thus indistinguishable from ordinary oscillator wander, and estimating it
separately is neither possible nor necessary.

To index the UT clock, we first define a common timeline. Let
\begin{equation}
  \label{eq:timeLine}
  \mathcal{T} = \{\tau_0, \tau_1, \dots, \tau_{N_z - 1} \}
\end{equation}
be the ordered set of UT clock times that correspond to the observation vector
in
(\ref{eq:obsVec}), and let $k(i,m)$ map the $m$th frame of the $i$th SV to its
time index, so that $\tau_{k(i,m)}$ is the UT TOA of the $m$th frame
of the $i$th SV.

We can then define the effective UT clock offset
\begin{equation}
  \label{eq:effectiveBias}
  b(k) \triangleq c\,\delta t_{\text{u}}(\tau_k) + \zeta(i(k)),
  \qquad
  \dot{b}(k) \triangleq c\,\delta \dot{t}_{\text{u}}(\tau_k)
\end{equation}
where $i(k)$ denotes the SV observed at the $k$th TOA, and we account for the
absorbed term in the process noise of Section~\ref{sec:clkProcess} so that the
reported covariance remains truthful.

Collecting the epoch-wise clock state $\vb{c}(k) \define [b(k) \,\,
\dot{b}(k)]\T$,
the complete state vector $\vb{x} \in \mathbb{R}^{n_x}$, with $n_x = 3 + 2N_z$,
is
\begin{equation}
  \label{eq:stateVec}
  \vb{x} = \begin{bmatrix}\vb{p}_{\text{u}}\T & \vb{c}\T(0) & \dots &
               \vb{c}\T(N_z-1)\end{bmatrix}\T
\end{equation}
where $\vb{p}_{\text{u}} \in \mathbb{R}^3$ is the UT's static Earth-centered,
Earth-fixed position.  We can then rewrite the predicted pseudorange as a
function
of the state vector in (\ref{eq:stateVec})
\begin{equation}
  \label{eq:predPseudo}
  \bar{\rho}(i,m) = \|\vb{p}_{\text{SV}}(i,m) - \vb{p}_{\text{u}}\| + b(k(i,m)) - c \dtfb(i,m)
\end{equation}
where $\vb{p}_{\text{SV}}(i,m)$ is the ephemeris-predicted position of the $i$th
SV at the
transmission time of the $m$th frame. Note that at the true $\vb{p}_{\text{u}}$,
the norm equals $r_{\text{u}} + e_{\text{u}}$,
which is what is shown in (\ref{eq:rhoPred}).

The corresponding row of the design matrix
$\mathbf{H}$ evaluated at the $m$th frame of the $i$th SV is the Jacobian of the
observation with respect to the state vector
\begin{equation}
  \label{eq:HrowSmoother}
  \mathbf{H}(i,m) = \begin{bmatrix}\hat{\vb{p}}\T(i,m) & \vb{s}\T\!\left(k(i,m)\right)
  \end{bmatrix}
\end{equation}
Here $\hat{\vb{p}}(i,m) = \partial \bar{\rho}(i,m) / \partial \vb{p}_{\text{u}}$
is
the line-of-sight unit vector from the SV to the UT, and the clock selector
$\vb{s}(k) \in \{0,1\}^{2N_z}$ is a unit vector selecting the bias entry of
$\vb{c}(k)$.  Each measurement thus observes the clock bias at its own epoch and
nothing else.

\subsection{Measurement Noise Covariance}
Ordinary least squares assumes independent and identically distributed
measurements.  Atmospheric delays, multipath, and time-varying SNR likely
introduce spatio-temporal correlations, which an optimal estimator must
characterize through a full measurement covariance matrix. An operational
estimator
may be able to perform nearly as well without the full matrix, assuming the
measurement noise is
approximately white.  We make no such assumption here and elect to derive the
full
matrix from our measurements.  Doing so allows us to (i) characterize the noise
covariance associated
with our measurements and (ii) calculate a more truthful formal estimation error
covariance that captures any measurement noise correlations.

Let $\vb{w}_\rho(i) = [w_\rho(i,0), \dots, w_\rho(i, |\mathcal{M}_i| - 1)]\T$
denote the
measurement noise vector for the $i$th SV over an FAI.  We model this vector as
a
multivariate Gaussian, $\vb{w}_\rho(i) \sim \mathcal{N}(\vb{0},
\mathbf{R}_\rho(i))$,
where the covariance matrix $\mathbf{R}_\rho(i)$ captures any intra-pass
correlations.  To develop an empirical model for $\mathbf{R}_\rho(i)$, we
compute the predicted
pseudoranges via (\ref{eq:predPseudo}) from the UT's surveyed ground-truth
position and the public
SV ephemeris, instead letting $b(k(i,m)) = c \delta t_\text{u}(i,m)$, where
$\delta t_\text{u}(i,m)$ is the true UT clock bias from simultaneous GPS
captures.
Subtracting the prediction from the measured pseudoranges isolates the combined
measurement noise and constant error term
\begin{equation}
  \hat{\vb{w}}_\rho(i) = \vb{\rho}(i) - \bar{\vb{\rho}}(i)
  = \vb{w}_\rho(i) + \zeta(i)\vb{1}
\end{equation}
where $\vb{1}$ is an all-ones vector of length $|\mathcal{M}_i|$.  Because
$\zeta(i)$ remains nearly constant over the brief FAI, subtracting the sample
mean from
the residuals isolates the zero-mean measurement noise
\begin{equation}
  \tilde{\vb{w}}_\rho(i) = \hat{\vb{w}}_\rho(i) - \left( \frac{1}{|\mathcal{M}_i|}
  \sum_{m=0}^{|\mathcal{M}_i| - 1} \hat{w}_\rho(i,m) \right) \vb{1}
\end{equation}
If the SNR is nearly constant over the FAI, the distribution of $\vb{w}_\rho(i)$
is
approximately stationary, which permits computing the biased empirical
autocorrelation of the detrended noise vector at the $d$th lag:
\begin{equation}
  r_{w_\rho}(i,d) = \frac{1}{|\mathcal{M}_i|} \sum_{m=0}^{|\mathcal{M}_i| - 1 - d}
  \tilde{w}_\rho(i,m) \tilde{w}_\rho(i,m+d)
\end{equation}
We construct $\mathbf{R}_\rho(i)$ as a symmetric Toeplitz matrix whose first row
is
the autocorrelation sequence $[r_{w_\rho}(i,0), r_{w_\rho}(i,1), \dots]$.
Enough time
passes between distinct SV passes to render their underlying noise processes
mutually uncorrelated, so the global pseudorange measurement covariance matrix
takes the block-diagonal form
\begin{equation}
  \label{eq:blockDiagR}
  \mathbf{R}_\rho = \mathrm{blkdiag}\big(\mathbf{R}_\rho(0), \mathbf{R}_\rho(1), \dots,
  \mathbf{R}_\rho(n-1)\big)
\end{equation}
The above measurement covariance matrix would be difficult for a UT to compute
operationally. Because the UT is assumed not to know its position, it would need
to fit its measurements to a polynomial model to extract residuals, which would
capture spurious correlation due to unmodeled orbital dynamics. Alternatively,
the UT could compute a diagonal measurement covariance matrix from its measured
SNR and the ZZB.

For the experiment presented in this work, the diagonal matrix would have been
acceptable. Applying the test from~\cite[Eq.~(7)]{stoica1977whiteness} to the
detrended residuals shows that each set of residuals is consistent with white
noise at a probability of false alarm of \SI{5}{\percent}. A test of the
estimator with both matrices yields differences in absolute solution errors and
variances of only a few centimeters and square centimeters, respectively.
\subsection{Clock Process Model}
\label{sec:clkProcess}
The temporal separation between passes allows the UT clock to drift, which
introduces significant time-dependent correlations between measurements.  We
model the UT clock offset as the sum of a Wiener process and an integrated
Wiener process~\cite{galleani2008tutorial}, driven by white frequency noise and
random-walk frequency noise with power spectral densities
\begin{equation}
  S_f = h_0/2, \qquad S_g = 2\pi^2 h_{-2} \nonumber
\end{equation}
respectively~\cite{brown2012introKf}, where the Allan variance coefficients
$h_0$ and $h_{-2}$ follow from the specific oscillator hardware.  Discretizing
this process over the interval $\Delta \tau_k = \tau_{k+1} - \tau_k$ between
consecutive epochs gives the standard two-state clock model
\begin{equation}
  \label{eq:clkDyn}
  \vb{c}(k+1) = \mathbf{F}(\Delta \tau_k)\,\vb{c}(k) + \vb{v}(k), \qquad
  \mathbf{F}(\Delta \tau) = \begin{bmatrix} 1 & \Delta \tau \\ 0 & 1 \end{bmatrix}
\end{equation}
with $\vb{v}(k) \sim \mathcal{N}\big(\vb{0}, \mathbf{Q}(\Delta \tau_k)\big)$.
Expressed in units of range, the entries of the process noise covariance are
\begin{equation}
  \label{eq:clkQentries}
  q_{11}(\Delta \tau) = c^2\!\left(S_f \Delta \tau + \frac{S_g \Delta
    \tau^3}{3}\right), \quad
  q_{12}(\Delta \tau) = \frac{c^2 S_g \Delta \tau^2}{2} \nonumber
\end{equation}
\begin{equation}
  \label{eq:clkQ}
  q_{22}(\Delta \tau) = c^2 S_g \Delta \tau, \qquad
  \mathbf{Q}(\Delta \tau) = \begin{bmatrix} q_{11} & q_{12} \\ q_{12} & q_{22}
  \end{bmatrix} \nonumber
\end{equation}

Absorbing $\zeta(i)$ into the clock states, as (\ref{eq:effectiveBias})
prescribes, inflates the process noise across a pass boundary.  If
$\zeta(i)$ is modeled as an independent zero-mean Gaussian with variance
$\sigma_{\zeta}^2$ for each SV, the increment $\zeta(i(k+1)) - \zeta(i(k))$
contributes a variance of $2\sigma_{\zeta}^2$ to the effective bias and nothing
to the drift.  The process noise covariance applied between epochs $k$ and $k+1$
is therefore
\begin{equation}
  \label{eq:clkQtilde}
  \tilde{\mathbf{Q}}(\Delta \tau_k) = \mathbf{Q}(\Delta \tau_k) +
  2\sigma_{\zeta}^2\, \vb{e}_1 \vb{e}_1\T \, \delta_k
\end{equation}
where $\vb{e}_1 = [1 \,\, 0]\T$ and $\delta_k$ equals one when epochs $k$ and
$k+1$ belong to different SVs and zero otherwise.  We set $\sigma_{\zeta}$ from
the CS clock estimation error reported above, which keeps the estimator's
covariance truthful without introducing an additional, unnecessary state.

We stack the process noise sequence as $\vb{V} = [\vb{v}\T(0) \, \dots \,
  \vb{v}\T(N_z-2)]\T$, with block-diagonal covariance $\mathbf{Q}_V =
  \mathrm{blkdiag}\big(\tilde{\mathbf{Q}}(\Delta \tau_0), \dots,
  \tilde{\mathbf{Q}}(\Delta \tau_{N_z-2})\big)$.  Rearranging (\ref{eq:clkDyn})
expresses each increment as a linear function of the state vector
\begin{equation}
  \label{eq:dynOperator}
  \vb{v}(k) = \vb{c}(k+1) - \mathbf{F}(\Delta \tau_k)\vb{c}(k)
  \triangleq \mathbf{A}_k \vb{x}, \qquad \vb{V} = \mathbf{A}\vb{x}
\end{equation}
where $\mathbf{A} = \left[\mathbf{A}\T_0 \, \dots \, \mathbf{A}\T_{N_z -
2}\right]\T \in \mathbb{R}^{2(N_z-1) \times n_x}$, with
$\mathbf{A}_k$ containing $-\mathbf{F}(\Delta \tau_k)$ and $\mathbf{I}_2$ in the
columns of $\vb{c}(k)$ and $\vb{c}(k+1)$ and zeros elsewhere.

\subsection{Batch Least-Squares Solution}
\label{sec:batchLS}
Let $\delta\vb{z}(\vb{x}) = \vb{z} - \bar{\vb{z}}(\vb{x})$ denote the
measurement residual formed from (\ref{eq:predPseudo}).  The estimator
minimizes the batch cost~\cite{y_barshalom01_tan}
  \begin{equation}
    \label{eq:batchCost}
    \begin{split}
      J(\vb{x}) = \tfrac{1}{2} & \delta\vb{z}\T(\vb{x})\, \mathbf{R}_\rho^{-1}
      \delta\vb{z}(\vb{x})                                                                         \\
                               & + \tfrac{1}{2} \left(\mathbf{A}\vb{x}\right)\T \mathbf{Q}_V^{-1}
      \left(\mathbf{A}\vb{x}\right)
    \end{split}
  \end{equation}
Linearizing $\delta\vb{z}$ about the current estimate and setting $\partial J /
  \partial \vb{x} = \vb{0}$ yields the Gauss--Newton update.  At the $j$th
iteration, with $\mathbf{H}$ assembled from the rows (\ref{eq:HrowSmoother})
evaluated at $\hat{\vb{x}}_{j-1}$,
\begin{equation}\label{eq:batchUpdate}
  \begin{split}
    \hat{\vb{x}}_j = \hat{\vb{x}}_{j-1} + \mathbf{\Lambda}^{-1}
    \big[ & \mathbf{H}\T \mathbf{R}_\rho^{-1} \delta\vb{z}_j                     \\
          & - \mathbf{A}\T \mathbf{Q}_V^{-1} \mathbf{A} \hat{\vb{x}}_{j-1} \big]
  \end{split}
\end{equation}
\begin{equation}\label{eq:infoMatrix}
  \mathbf{\Lambda} = \mathbf{H}\T \mathbf{R}_\rho^{-1} \mathbf{H} +
  \mathbf{A}\T \mathbf{Q}_V^{-1} \mathbf{A}
\end{equation}

We initialize the estimator with $\hat{\vb{x}}_0 = [\vb{p}_0\T, \vb{0}\T]\T$,
where
$\vb{p}_0$ is a coarse initial position guess and the $2N_z$ clock states are
zeroed.  The clock states enter (\ref{eq:predPseudo}) linearly, so the
range to each SV remains the only nonlinearity and a coarse clock initialization
costs nothing.  The UT iterates (\ref{eq:batchUpdate}) until the position
correction norm falls below a specified convergence threshold or a predetermined
number of iterations is reached, whichever comes first.  Upon convergence,
the estimator yields the final state $\hat{\vb{x}}$ alongside the covariance
matrix
\begin{equation}
  \label{eq:finalSolCovar}
  \mathbf{P}_{\hat{\vb{x}}} = \mathbf{\Lambda}^{-1}
\end{equation}
\section{Signal Capture}
\label{sec:sigCapture}
To emulate the fused LEO GNSS concept with the existing Starlink network, we
developed a dual-tier signal capture framework that contrasts a high-fidelity CS
architecture with a degraded, consumer-grade UT architecture. This section
describes each architecture in turn.

\subsection{CS Capture Architecture}
\begin{figure*}[t]
  \centering
  \tikzset{
  rightarrow/.style={-{Latex[length=3mm, width=2mm]}, line width=1pt,
  join=round},
  dashed_ln/.style={-, dashed, line width=1pt, join=round},
  dashed_ln_rightarrow/.style={-{Latex[length=3mm, width=2mm]}, dashed, line
  width=1pt, join=round}
  }
  \begin{adjustbox}{max width=\textwidth}
    \input{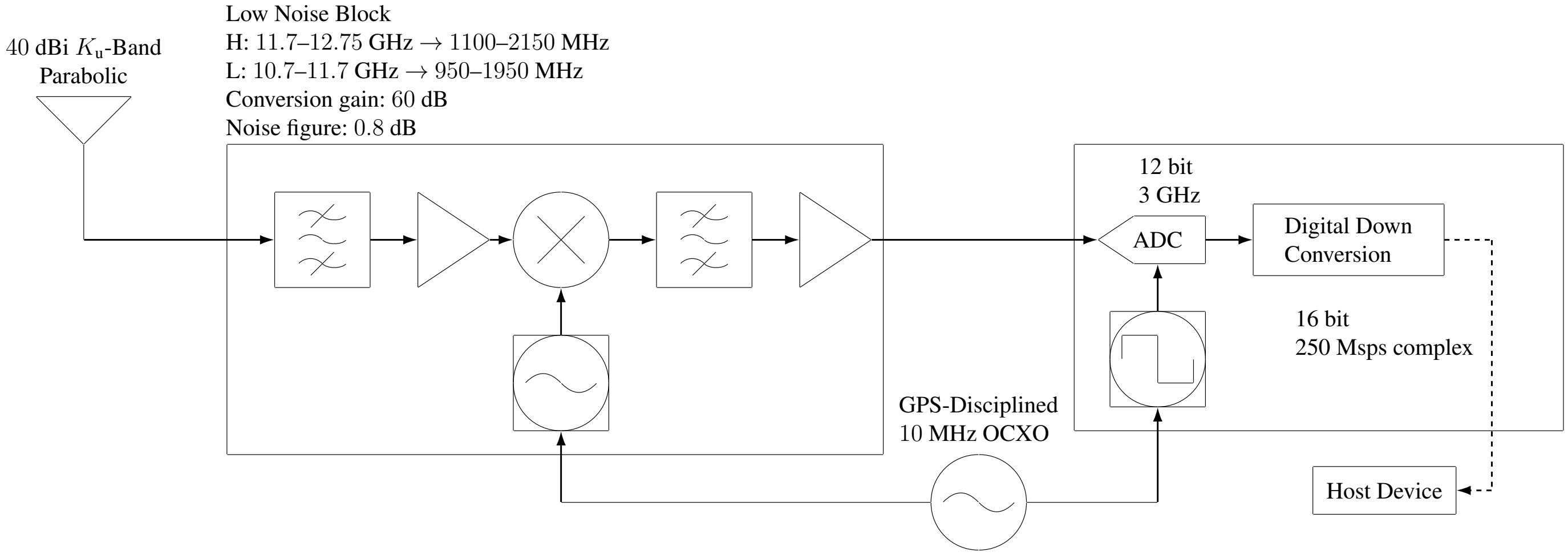}
  \end{adjustbox}
  \caption{Block diagram of the CS signal capture process.}
  \label{fig:capture_block_diagram}
\end{figure*}

The CS must extract high-quality TOA measurements and decode frame data to
provide the UT with accurate frame clock offsets $\dtfb(i,m)$ and pilot
sequences. Achieving the necessary post-correlation SNR requires a highly
directional antenna and a wideband radio front end.

As illustrated in Fig.~\ref{fig:capture_block_diagram}, the CS capture hardware,
located on the University of Texas at Austin campus with a clear view of the
sky, comprises four main components. Its antenna is a steerable
\SI{90}{\centi\meter} offset parabolic dish with an approximate \SI{3}{\degree}
beamwidth, yielding a nominal gain of \SI{40}{\dBi} at \SI{12.5}{\giga\hertz},
less \SIrange{4}{5}{\decibel} of polarization and feedhorn misalignment losses.
The dish is actively steered according to public TLEs. A low-noise block
downconverter with a \SI{60}{\decibel} conversion gain and a \SI{0.8}{\decibel}
noise figure converts the \SIrange{10.7}{11.7}{\giga\hertz} and
\SIrange{11.7}{12.75}{\giga\hertz} Ku bands to intermediate frequencies of
\SIrange{950}{1950}{\mega\hertz} and \SIrange{1100}{2150}{\mega\hertz},
respectively. A dual-channel software-defined radio then samples the signal
continuously at \SI{250}{\mega\hertz} with 16-bit complex resolution, providing
roughly \SI{200}{\mega\hertz} of usable bandwidth, enough to capture the bulk of
Starlink's $\Fs = \SI{240}{\mega\hertz}$ downlink signal. Finally, an OCXO
drives all sampling. Because of the narrow \SI{3}{\degree} beamwidth, the CS can
track and capture only a single SV at a time.

\subsection{UT Capture Architecture}

We model the UT as a consumer-grade device with a low-gain antenna and a narrow
front-end bandwidth. The UT capture hardware mirrors the block diagram in
Fig.~\ref{fig:capture_block_diagram} but with two intentional downgrades. First,
a compact \SI{80}{\milli\meter}-diameter Ku-band feedhorn replaces the
\SI{90}{\centi\meter} parabolic dish, significantly reducing the raw antenna
gain. Second, the sampling rate is decimated to \SI{25}{\mega\hertz} (16-bit
complex), reducing the captured bandwidth and, therefore, the precision of the
extracted TOAs.

The capture hardware imposes one further limitation. Whereas an operational
Starlink UT has an agile RF front end capable of simultaneously downmixing
signals from multiple SVs across different center frequencies, our experimental
capture hardware has only a single RF processing chain. Consequently, both the
CS and the UT can observe only one SV at a time, leaving idle periods of several
minutes between SV passes.

In a true consumer device, a free-running temperature-compensated crystal
oscillator would drift unpredictably during these long gaps, breaking the
continuity of the clock bias estimation. To bridge these artificial,
hardware-induced idle periods, we drove the experimental UT's sampling clock
with an OCXO.

\section{Experimental Results}
\label{sec:experiment}
This section describes the experimental scenario and then evaluates the PNT
performance obtained with three TOA extraction methods.

\subsection{Scenario Description}
Starlink signal data were recorded over a 13-minute period in January 2025. The
UT captured data continuously to maintain local clock coherence across the
entire interval, while the CS captured 60-second bursts as individual SVs became
observable.

\begin{figure}[t]
  \centering
  \includegraphics[width=\columnwidth]{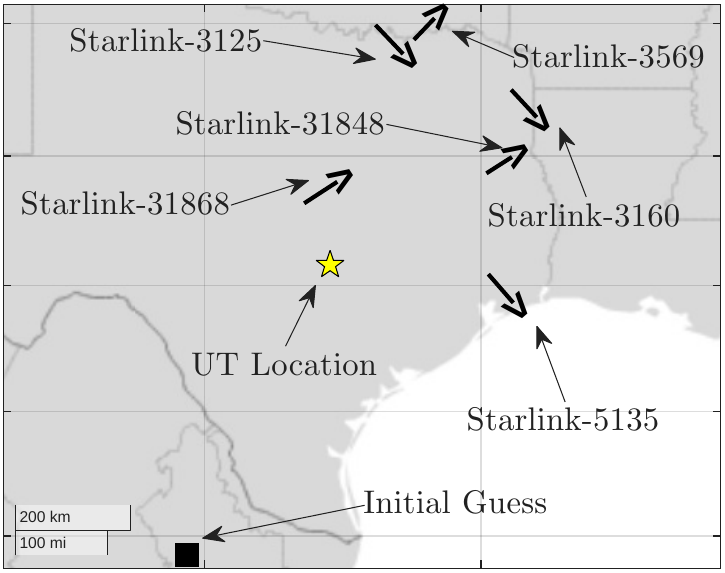}
  \caption{The ground tracks of the recorded Starlink SVs. In total, six SVs
  were recorded. In order of capture, their identifiers are 5135, 31868, 3160,
  3569, 3125, and 31848. The yellow star marks the location of the UT, and the
  black square marks the location of the WNBLS initialization. Arrows indicate
  the direction of travel.}
  \label{fig:groundTracks}
\end{figure}

Both stations monitored a center frequency of \SI{12.075}{\giga\hertz},
corresponding to channel 6 in~\cite{humphreys2023starlinkSignalStructure}. Over
the 13-minute window, we tracked six distinct SVs. While an operational UT
equipped with an agile phased-array antenna would observe significantly more SVs
in this timeframe, our experimental capture was limited by the mechanical slew
rate of the parabolic dish and the uncertainty of active transmissions on the
targeted channel. Fig.~\ref{fig:groundTracks} shows the ground tracks of the
recorded Starlink SVs.

For each of the six SVs, measurements were aggregated over a \SI{10}{\second}
window of a single FAI. The idle periods between successive SV captures ranged
from \SIrange{100}{280}{\second}.

At the UT, we measured the TOAs $\ttu(i,m)$ by three methods for comparison. The
nearest-sample (NS) method forms local replicas from fully decoded frames and
estimates TOA by coarse, nearest-sample correlation, analogous to standard GNSS
acquisition. Importantly, the NS method first upsamples the received data to the
full \SI{240}{\mega\hertz} bandwidth before correlation, providing more accuracy
than a downsampled \SI{25}{\mega\hertz} local replica would. The
super-resolution (SR) method forms the same local replicas but estimates TOA
with the maximum-likelihood techniques detailed
in~\cite{qin2025toaDopplerPlans}. The optimal-pilots (OP) method relies solely
on the ZZB-optimal pilot allocations described in Section~\ref{sec:dataAiding},
again with maximum-likelihood TOA estimation.

At the CS, we measured the reference TOAs $\ttr(i,m)$ exclusively with
full-frame maximum-likelihood techniques~\cite{qin2025toaDopplerPlans} and from
them computed the frame clock offset $\dtfb(i,m)$. This offset was then
forwarded to the UT to predict pseudorange measurements according to
(\ref{eq:rhoPred}).

To evaluate the estimator's ability to recover the clock bias, we artificially
injected a known initial clock bias by initializing $\ttu(0,0)$ as
\begin{equation}
  \ttu(0,0) = \tf(0,0) + \hatdTOF(0,0)
\end{equation}
where $\hatdTOF(0,0)$ is a simulated, coarse initial guess of the true time of
flight $\dTOF(0,0)$. The injected initial clock bias is therefore
\begin{equation}
  \delta\tu(0,0) = \hatdTOF(0,0) - \dTOF(0,0)
\end{equation}
Subsequent estimates of clock bias can be compared to the known initial value
plus the known change in clock bias as determined by the simultaneous GPS
capture.

The 13-minute recording yielded \SI{60}{\second} of active SV observation time.
Processing this dataset produced $N_z = 2956$ valid $\ttu(i,m)$ measurements
with fully decoded frames (NS and SR) but only $N_z = 2637$ measurements with OP
local replicas. This discrepancy arises because the OP method lacks the
extensive processing gain provided by fully decoded frames, precluding the
detection of weaker signals. Measurements from a single SV occur at arbitrary
multiples of the frame period $\Tf$.

\subsection{PNT Performance}
The WNBLS estimator was initialized as detailed in Section~\ref{sec:estimation}
with an initial position guess in Monterrey, Mexico (shown in
Fig.~\ref{fig:groundTracks}), approximately \SI{572}{\kilo\meter} away from the
true UT location.
An OCXO drove the UT receiver in this experiment. The clock parameters were $h_0
= 3.8\times 10^{-22}$ and $h_{-2} = 2.3\times 10^{-24}$.

\begin{figure}[t]
  \centering
  \includegraphics[width=\columnwidth]{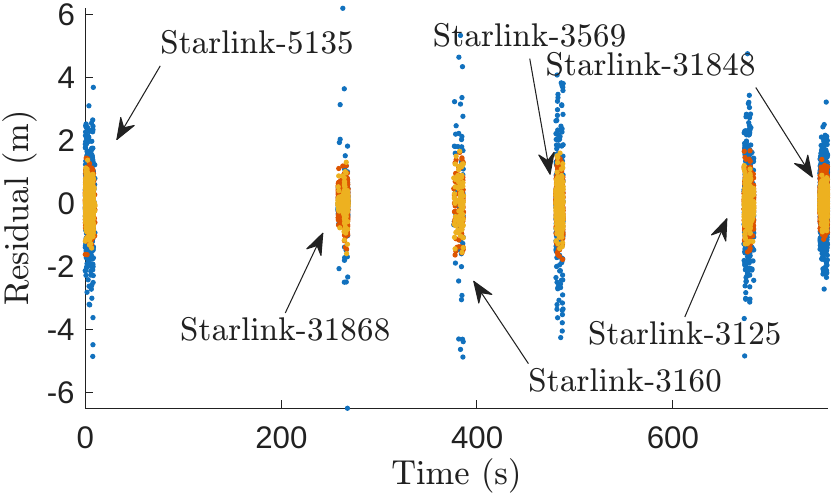}
  \includegraphics[width=\columnwidth]{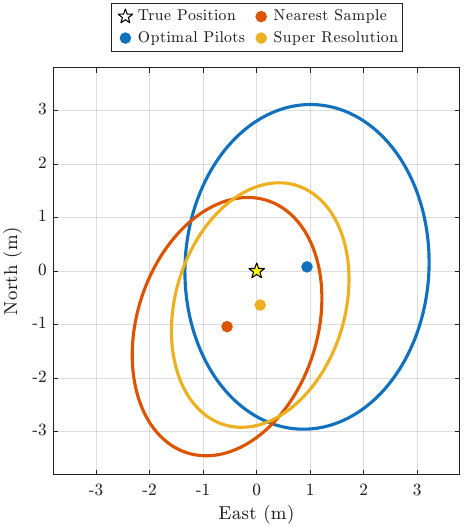}
  \caption{The post-fit WNBLS residuals (top) for the different TOA extraction
  methods, highlighting RMSE differences and long observation gaps. The bottom
  plot shows the corresponding 2D position solutions and formal
  \SI{95}{\percent} ellipses alongside the true UT location (yellow star).}
  \label{fig:solution}
\end{figure}

\begin{figure}[t]
  \centering
  \includegraphics[width=\columnwidth]{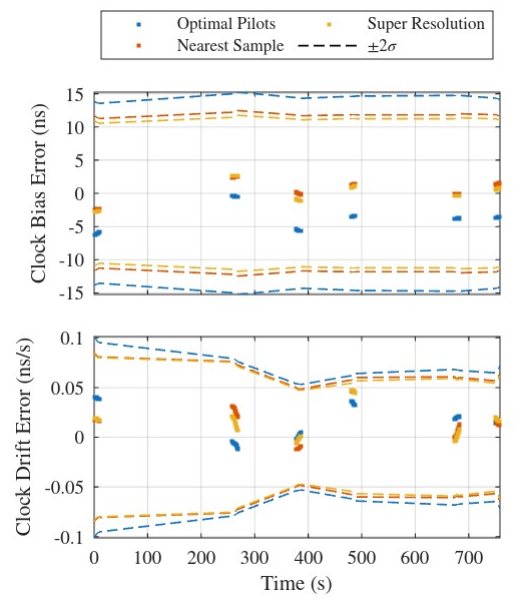}
  \caption{Clock bias and drift estimates and their associated $2\sigma$ bounds.
  Estimates of clock bias and drift occur only when measurements are available.}
  \label{fig:timeSolution}
\end{figure}

The WNBLS position solutions and corresponding post-fit residuals for the OP,
NS, and SR TOA extraction methods are plotted in Fig.~\ref{fig:solution}. The
residuals are nearly zero-mean across all methods, and the gaps between them
show the extended idle periods between SV observations. Timing results are shown
in Fig.~\ref{fig:timeSolution}.

\begin{table*}[t]
  \caption{Positioning performance for different TOA extraction methods}
  \label{tab:posResults}
  \centering
  \small
  \sisetup{
    table-number-alignment = center,
    round-mode = places,
    round-precision = 2
  }
  \begin{tabular}{l
      S[table-format=1.2]
      S[table-format=1.2]
      S[table-format=1.2]
      S[table-format=1.2]
      S[table-format=1.2]
      S[table-format=1.2]}
    \toprule
    \multirow{2}{*}{Method}
       & {Avg. Res. RMSE}

       & \multicolumn{3}{c}{Horizontal Err. (m)}

       & \multicolumn{2}{c}{Vertical Err. (m)}                                      \\
    \cmidrule(lr){3-5} \cmidrule(lr){6-7}
       & {(m)}

       & {2D}

       & {95\% Semi-Maj.}

       & {95\% Semi-Min.}

       & {Up}

       & {95\%}                                                                     \\
    \midrule
    OP & 1.09                                    & 0.94 & 3.03 & 2.28 & 1.23 & 4.35 \\
    NS & 0.51                                    & 1.17 & 2.47 & 1.69 & 0.31 & 3.62 \\
    SR & 0.38                                    & 0.63 & 2.33 & 1.58 & 0.78 & 3.37 \\
    \bottomrule
  \end{tabular}
\end{table*}

\begin{table}[t]
  \caption{Timing performance for different TOA extraction methods}
  \label{tab:timeResults}
  \centering
  \small
  \sisetup{
    table-number-alignment = center,
    round-mode = places,
    round-precision = 2
  }
  \begin{tabular}{l
      S[table-format=1.2]
      S[table-format=2.2]
      S[table-format=1.4, round-precision=4]
      S[table-format=1.4, round-precision=4]}
    \toprule
    {Method}                   &
    {$\delta \tu$}             &
    {$\delta \tu$}             &
    {$\delta\dot{t}_\text{u}$} &
    {$\delta\dot{t}_\text{u}$}                                  \\
    {}                         &
    {RMSE}                     &
    {Avg. 95\% Err.}           &
    {RMSE}                     &
    {Avg. 95\% Err.}                                            \\
    {}                         &
    {(ns)}                     &
    {(ns)}                     &
    {(ns/s)}                   &
    {(ns/s)}                                                    \\
    \midrule
    OP                         & 3.85 & 14.45 & 0.0229 & 0.0744 \\
    NS                         & 1.82 & 11.86 & 0.0221 & 0.0669 \\
    SR                         & 1.79 & 11.19 & 0.0207 & 0.0649 \\
    \bottomrule
  \end{tabular}
\end{table}

Tables~\ref{tab:posResults} and~\ref{tab:timeResults} summarize the positioning
and timing performance, respectively. Of the three methods, SR yields the
tightest \SI{95}{\percent} bounds in every category: a horizontal error ellipse
with semi-major and semi-minor axes of \SI{2.33}{\meter} and \SI{1.58}{\meter},
a vertical error bound of \SI{3.37}{\meter}, and an average timing error bound
of \SI{11.19}{\nano\second}. For every method, the actual position and timing
errors fell well within the corresponding \SI{95}{\percent} bounds.

Both the SR and NS methods leverage fully decoded frames, providing roughly
\SI{45}{\decibel} of processing gain (calculated as
in~\cite{qin2026starlinktemplate}). By contrast, the OP method operates with a
lower processing gain of approximately \SI{35}{\decibel}. Despite this handicap,
the OP method still yields a viable navigation solution at a fraction of the
data processing overhead. However, for a UT possessing the necessary
computational resources to decode the data payload, the SR method is preferable,
maximizing accuracy without imposing additional capacity costs.

\section{Conclusion}
\label{sec:conclusion}
A mock implementation of a fused LEO GNSS based on Starlink's Ku-band downlink
has been developed and evaluated experimentally. A high-fidelity control segment
(CS) receiver derived frame clock corrections and pilot data and relayed them to
a user terminal (UT), emulating the models that a constellation operator would
broadcast. Despite the UT's degraded, consumer-grade receiver architecture with
a narrow-bandwidth front end and sparse SV observations, super-resolution (SR)
TOA extraction yielded \SI{95}{\percent} error bounds of \SI{2.33}{\meter}
horizontally (semi-major axis), \SI{3.37}{\meter} vertically, and
\SI{11.19}{\nano\second} in timing, and the actual solution errors fell well
within these bounds.

These results indicate that an accurate fused LEO GNSS service could be
implemented over the existing Starlink infrastructure if SpaceX can monitor SV
ephemerides to within a few meters and SV clocks to within
\SI{10}{\nano\second}. Alternatively, a third-party service could achieve
comparable performance by deploying an independent network of reference stations
at the service-cell level to broadcast real-time timing and ephemeris
corrections to users.

In either deployment scenario, this work has established the potential viability
of pseudorange-based LEO PNT as a robust, resilient backup to legacy GNSS. In
the near term, the presented results encourage further research in pseudorange
PNT with Starlink. In particular, the use of wide-beamwidth (e.g., feedhorn)
antennas and/or multichannel receivers at both the CS and the UT could improve
SV geometry by allowing more SVs to be observed at once. Looking further ahead,
as LEO constellations continue to densify and transmit wider bandwidths, we
expect LEO PNT to eventually provide navigation and timing services that are
fundamentally superior to those of legacy GNSS.

\section*{Acknowledgment}
Research was supported by the U.S. Department of Transportation under Grant
69A3552348327 for the CARMEN+ University Transportation Center, and by
affiliates of the 6G@UT center within the Wireless Networking and Communications
Group at The University of Texas at Austin.  The authors used Claude (Opus 5.5,
Anthropic) to copy-edit the manuscript for grammar, spelling, and prose as per
an in-house style guide. It was also used to generate
Fig.~\ref{fig:fusedLEOConops} from a drawing provided by the authors. The
authors reviewed and take responsibility for all content.

\bibliographystyle{IEEEtran}
\bibliography{./pangea}

\end{document}